\documentclass[nofootinbib,
amsmath,amssymb,
 aps,
prb,
twocolumn,
]{revtex4-2}\usepackage[english]{babel}
\usepackage[utf8]{inputenc}
\usepackage[colorinlistoftodos, color=green!40, prependcaption]{todonotes}
\usepackage[colorlinks=true,
            linkcolor=red,
            citecolor=blue,
            urlcolor=blue]{hyperref}

\usepackage{xcolor}
\usepackage{amsthm}
\usepackage{mathtools}
\usepackage{physics}
\usepackage{graphicx}

\usepackage[normalem]{ulem}
\usepackage{soul}
\usepackage{pgfplots}
\newcommand{\mbf}[1]{\mathbf{#1}}
\newcommand{\fig}[1]{Fig.~\ref{#1}}
\newcommand{\Fig}[1]{Figure~\ref{#1}}

\newcommand{\Sec}[1]{Sec.~\ref{#1}}

\newcommand{\paras}[2]{paragraphs~\ref{#1},~\ref{#2}}

\newcommand{\eqn}[1]{Eq.~(\ref{#1})}

\definecolor{orange}{rgb}{1,0.5,0}

\definecolor{pink}{HTML}{f52c86}

\newcommand{\mycite}[1]{\cite{#1}}

\newcommand{\app}[1]{app.~\ref{#1}}

\newcommand{\bl}[1]{\boldsymbol{#1}}

\def \veps {\varepsilon}
\newcommand{\ignore}[1]{}

\def\k {{\mbf{k}}}

\ifdefined\added
\else
\newcommand{\added}[2][]{\textcolor{blue}{#2}\textsuperscript{\small\textcolor{red}{#1}}}
\definecolor{shadecolor}{RGB}{80,100,80}
\definecolor{pink}{RGB}{220,100,100}
\usepackage{marginnote}
\newcounter{mparcnt}

\fi
\newcommand\minus{\setbox0=\hbox{-}\vcenter{\hrule width\wd0 height \the\fontdimen8\textfont3}}

\def\mcH {{\mathcal{H}}}
\def\Hk{{\mcH(\mathbf{k})}}
\def \Do{{D^{(0)}_{xy}}}
\def\Df{D^{(1)}_{xy}}
\def\Dex{D^{(\text{ext})}_{xy}}

\newcommand{\nn}[2]{\langle #1,#2\rangle}
\def\aTIII{{\alpha\,\minus{\cal T}^3}}

\DeclareMathSymbol{\shortminus}{\mathbin}{AMSa}{"39}
\newcommand{\sminus}{\scalebox{0.75}[1.0]{\( - \)}}
\ifdefined\redacton
    \ifdefined\indicateredacted
        \newcommand{\redacted}[1]{\textcolor{gray!70}{#1}}
        
    \else
        \newcommand{\redacted}[1]{}
        
    \fi
\else
        \newcommand{\redacted}[1]{#1}
        
\fi

\begin{document}
\title{Spin Injection Route to Magnon Berry Curvature Dipole}
\author{Atul Rathor}
\email[]{atulrathor@bose.res.in}\affiliation{S. N. Bose National Centre for Basic Sciences,
JD Block, Sector-III, Salt Lake City, Kolkata - 700 106, India}
\ifdefined\twoauthors\else
\author{Saurav Kantha}
\affiliation{S. N. Bose National Centre for Basic Sciences,
JD Block, Sector-III, Salt Lake City, Kolkata - 700 106, India}
\fi
\author{Arijit Haldar}
\email[]{arijit.haldar@bose.res.in}

\affiliation{S. N. Bose National Centre for Basic Sciences,
JD Block, Sector-III, Salt Lake City, Kolkata - 700 106, India}

\date{\today} 

\begin{abstract}
Berry curvature of Bloch bands arising in lattice systems can induce a Hall response even in the absence of topology due to the so-called Berry-curvature dipole (BCD). Such a response is universal and, in principle, should occur as a thermal-Hall effect in magnon systems under the application of a temperature gradient. However, this effect intrinsically appears as a non-linear (second-order) response to the temperature gradient 
making experimental detection difficult. Here, we propose an alternate route to access BCD in magnons. 
By utilizing the process of spin-injection  in conjunction with a temperature gradient, 
we uncover two previously unreported contributions to the BCD-induced Hall response for magnons -- one that is linear in temperature gradient, and the other is non-linear in the magnon-chemical potential gradient arising from spin injection.
As an added benefit of our approach, both these responses extract distinct moments of the genuine BCD distribution over the magnon bands, as opposed to the recently reported extended BCD in magnons.
We use  Boltzmann transport theory to derive the expression for the magnon-Hall response in the presence of a thermal gradient and spin injection. Furthermore, using this expression, we offer predictions for the BCD-induced magnon-Hall effect to be observed in experiments for ferro, anti-ferro and ferri magnetically-ordered models on various lattices, including the honeycomb lattice, the kagome lattice, and the dice lattice.
\end{abstract}

\maketitle

\section{Introduction} \label{sec:intro}

Magnon\cite{bloch_zur_1930}-based transport phenomena are at the heart of spintronics and magnonics\cite{chumak2015magnon}, where spins or their collective excitations, such as magnons in quantum magnets, are leveraged to provide energy-efficient transfer and storage of information.
Unlike electrons,  magnons do not carry an electric charge, and hence, magnon currents are immune to energy losses due to joule heating. The absence of joule heating leads to a 
dramatic increase in energy-utilization efficiency while performing operations such as information transfer\cite{vonmeier}.
Thus, making magnonics, and in general spintronics, a lucrative alternative to conventional electronics where devices based on electron charge are the norm. 

Multiple spin-based alternates to traditional charge-based devices have  been reported, and novel spin devices without any charge-based counterparts, including spin valves\cite{jansen2003spin}, spin diodes\cite{flatte2001unipolar,eisenstein2017quantum}, etc., have also been engineered. There have also been proposals and demonstrations of various devices utilizing magnons, including magnetic racetrack memories and others \cite{banerjee_advances_2023,behin-aein_proposal_2010,cao_prospect_2020,chumak_magnon_2015,fong_spin-transfer_2016,johnson_magnetoelectronic_2000,kroutvar_optically_2004,parkin_magnetic_2008,patra_all-spin_2018,raman_interface-engineered_2013}. Therefore, exploring magnon transport and understanding their similarities and distinctions with electronic phenomena is vital from the point of view of both fundamental and applied research.

Although electrons and magnons are intrinsically distinct, with differing origins and properties such as electric charge, their response in lattice systems share notable similarities. These similarities are a consequence of the geometric properties of the underlying Bloch bands; the latter naturally arise due to the periodicity inherent in lattice systems. Specifically, the geometry induced by Bloch bands on quantum states can be quantified by the so-called quantum geometric tensor \cite{cayssol_topological_2021,ma_abelian_2010,gianfrate_measurement_2020-1,wei_quantum_2023}, whose real part, dubbed the quantum metric tensor, describes the distance between states, and the imaginary part gives the Berry curvature\cite{berry1984quantal} that encodes the effects of an emergent gauge field. In this sense, the constraints set by the geometry of Bloch bands may be considered universal, and any quasiparticle, either electrons or magnons, whose dispersions form bands, are subject to the rules of band geometry.

A powerful and remarkable consequence of Berry curvature is connecting geometry with topology, exemplified by topological invariants, such as Chern-number, that can be mathematically expressed in terms of the Berry curvature\cite{kohmoto1985,PhysRevLett.51.2167}. Having a non-zero value for such topological invariants results in quantized Hall-type conductance that may be seen in transport experiments.
Following the geometry-topology connection, research into topology arising from Bloch bands has been extensive. Several breakthroughs in this area, such as topological insulators, spin-Hall effect, etc., have been presented\cite{xiao_berry_2010} and soon generalized to other platforms such as superconductors\cite{liu2018majorana, zhu2018tunable, laubscher2019fractional,wang2018high}, photonics\cite{el2019corner}, etc\cite{peterson2018quantized,imhof2018topolectrical}. Topological phases using magnons in lattice spin systems were also discovered\cite{onose2010observation,diaz2019topological,zhang2013topological,mook2014edge,cai2019observation,malz2019topological,mochizuki2014thermally,zhu2021topological,owerre2018photoinduced,mook2021chiral} along these lines. Even higher-order generalizations of topological insulating phases, dubbed Higher-order topology, have been presented for electronic\cite{benalcazar2017science, schindler2018higher} as well as spin systems\cite{HirosawaPRL2020, diaz2019topological, Haldar.Higher.PRB} and experimentally observed for electrons \cite{aggarwal2021evidence,schindler2018higher}. 

It is therefore natural to ask whether Berry-curvature ($\Omega(\k)$), defined for Bloch momentum $\k$ of the first Brillouin zone (BZ), can lead to geometry-induced transport phenomena beyond band topological effects. Indeed,  for electrons, \cite{sodemannQuantumNonlinearHall2015} showed that Berry-Curvature Dipole
\begin{equation}
\text{BCD}=
\int_{\text{\tiny BZ}} \; f_0\;\partial_{\k}\boldsymbol {\Omega} (\textbf{k})\label{eq:BCD_electron}
\end{equation}
defined as an integral over the BCD-distribution $\partial_{\k}\boldsymbol {\Omega} (\textbf{k})$, a \emph{purely geometric quantity},  weighted by the Fermi-Dirac function $f_0$, can produce experimental signatures similar to bands described by a non-zero Chern number even in the absence of topology. In particular, a Bloch band with a finite BCD but zero Chern number will generate a quantum Hall response transverse to an applied electric field, just like a topological band carrying a non-zero Chern number. Recently, BCD-induced Hall response  have also been predicted to occur in magnonic systems lacking topological properties\cite{kondoNonlinearMagnonSpin2022}.

Like the topological magnon-Hall response, BCD-induced magnon Hall response is also proposed to occur as a thermal Hall effect, in which, a transverse spin current develops in response to an applied temperature gradient instead of an electric field. However, there are significant differences between BCD and Chern number-induced thermal Hall responses. First, the Hall conductance from BCD is not quantized as the topological Hall effect. Second, while the topological thermal   Hall effect depends linearly on the applied temperature gradient, the BCD-Hall response occurs as a non-linear second-order effect. The latter makes the BCD-induced Hall effect inherently difficult to be measured in an experimental setup compared to the topological Hall effect. In this paper, we bring the thermal Hall response in magnonic systems, arising from BCD, on equal footing with the topological Hall effect by proposing a setup to measure BCD-Hall conductance in the linear order of temperature gradient.

The key ingredient in our proposal, designed for electrically insulating magnetic materials,
is to set up a pre-existing magnon current in the material sample independent of the temperature
gradient to be applied.
Interestingly, such an initial magnon current can be set up using spin injection 
(described in detail later), a standard procedure in spintronic experiments \cite{cornelissenMagnonSpinTransport2016}. 
In this process, a metal with spin-orbit coupling is placed in contact with the magnetic sample. 
A charge current driven through the metal generates a  spin current, perpendicular to the charge current, via the spin 
Hall effect, which diffuses through the interface into the magnet. As the magnet is an electrical 
insulator, the injected spin current naturally manifests as magnon current inside the sample.
We show that in such a scenario the application of the temperature gradient couples to the
pre-existing magnon-current, and modifies it, to produce a transverse Hall current proportional to the temperature gradient. 
Overall, our work introduces three significant advances to the area of BCD-induced magnon transport.
First, we bring a second-order magnon spin current to the linear order in the applied temperature gradient, hence making the experimental measurement of the transverse magnon current more accessible. 
Secondly, as we show , the transverse current generated in our proposal is a measure of the true BCD of magnon bands, a purely geometrical quantity, in contrast to a BCD-like quantity called Extended BCD, which was proposed in an earlier work \cite{kondoNonlinearMagnonSpin2022}. Thus, our work complements \cite{kondoNonlinearMagnonSpin2022}, and generalizes the BCD-induced magnon-Hall response beyond the application of temperature gradients. Third, we also discover an additional response that couples to the BCD-distribution in the BZ and which is non-linear in the magnon-chemical potential gradient induced by the spin-injection process.

This paper is organized as follows. In section \ref{sec:proposal}, we describe the details of our proposal along with the experimental setup. In section \ref{sec:BTE}, we  use Boltzmann transport formalism to  obtain analytical expression for magnon Hall current, arising from the simultaneous application of temperature and magnon chemical potential gradients, and reveal the contributions of the BCDs to the current. Next, we develop the spin-wave theory for ferro, anti-ferro and ferri-magnetic models on several topical {and} experimentally {relevant} lattice geometries in section \ref{sec:Models}.  We then present our results and predictions for the BCDs and extended BCD for these models  in section \ref{sec:Results}. We summarize our work and discuss the implications of our proposal in section \ref{sec:summary}. \section{Proposed Setup}\label{sec:proposal}
\begin{figure}
    \centering
\includegraphics[width=0.49\textwidth]{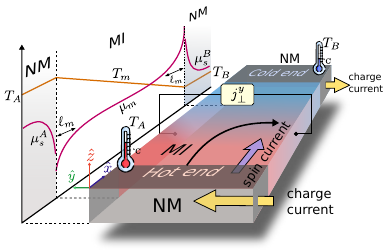}
    \caption{Proposed setup for measuring transverse magnon-spin current ($j^{y}_\perp$) generated from spin-injection and applied temperature-gradient: A magnetic insulator (MI) is placed between two non-magnetic metals (NM) at ends $A$ and $B$, maintained at temperatures $T_A$ and $T_B$. Charge current $j_c^y$, applied to $A$ along $\hat{\textbf{y}}$, produces a non-zero spin accumulation $\mu_s^A$ in the metal and a non-zero magnon-chemical potential $\mu_m$ inside the MI via the spin-Hall effect. A non-zero $\mu_s^B$ is set up in the metal at the other end $B$ via the inverse spin-Hall effect. Typical spatial profiles for $\mu_s^A$, $\mu_s^B$, $\mu_m$ and the magnon temperature $T_m$  are shown in the projected XZ plane (left). Symbol $l_m$ indicates the magnon diffusion length.} 
    \label{fig:setup}
\end{figure}

{Our setup comprises of an electrically insulating  film of a magnetically-ordered material (or sample) having dimensions $L_x\times L_y$ and placed on the $x$-$y$ plane as shown in \Fig{fig:setup}.}{}
It is connected to two metallic leads having spin-orbit coupling, at $x~=~-L_x\slash2$ (end A) and $x=L_x/2$ (end B) maintained at temperatures $T_A$ and $T_B$ respectively with $T_A>T_B$. Due to local thermal equilibrium, the temperature of magnons in the sample at either end is expected to be the same as the leads.

To achieve spin injection into the magnetic sample, a charge current $j_c^y$ is driven along the negative $\hat{\textbf{y}}$ axis in the metal at end A using an electric field $\mathcal{E}$. Due to the presence of spin-orbit coupling, this charge current $j_c^y$ generates a spin current $j_s^x$ inside the metal along the x-axis via the spin Hall effect. Both the spin and charge currents in the metal occur in the diffusive regime and can be described by the following  equations respectively \cite{duineSpintronicsMagnonBoseEinstein2015,cornelissenMagnonSpinTransport2016}
\begin{subequations}\label{eq:j_metal}
\begin{align}
     j_c^y  &= \sigma \mathcal{E} + \frac{\sigma_{SH}}{2e} \frac{\partial \mu_s^A}{\partial x} \label{eq:jc}    \\
    j_s^x  = &-\frac{\sigma\hbar}{4e^2}\frac{\partial \mu_s^A}{\partial x}-\frac{\hbar\sigma_{SH} \mathcal{E}}{2e}.
    \label{eq:js}
\end{align}
\end{subequations}

Here, $\sigma$ is the charge conductivity of the metal, $e$ is the charge of the electron, and $\sigma_{SH}$ is the spin-Hall conductivity. The symbol  $\mu_s^A$ denotes spin accumulation which is the difference between the chemical potentials of spin-up and spin-down electrons in the metal.  The spin accumulation $\mu_s^A$ inside the metal is non-zero under non-equilibrium conditions, such as when charge currents are driven through the metal, and since the rate of thermalization of electrons is lower than the rate of spin-flip relaxation\cite{cornelissenMagnonSpinTransport2016,duineSpintronicsMagnonBoseEinstein2015}. These effects combined lead to an effective description of the up and down spin electrons via separate Fermi-Dirac distributions characterized by distinct temperatures and chemical potentials, consistent with a non-zero $\mu_s^A$
\cite{cornelissenMagnonSpinTransport2016,duineSpintronicsMagnonBoseEinstein2015}. To account for the spin injection process theoretically, it will be sufficient to focus on the difference in chemical potentials of the two spins, i.e. $\mu_s^A$. \eqn{eq:js} is  supplemented by the equation,
\begin{equation}
    \frac{\partial j_s^x}{\partial x}=-\Gamma_s \mu_s^A,
    \label{eq:js_ph}
\end{equation}
that describes the spin-flip relaxation. The phenomenological constant $\Gamma_s$  contains material-specific information regarding the spin-flip relaxation process and, in principle, depends on the density of states of electrons at the Fermi level inside the metal. Combining \eqn{eq:js} and \eqn{eq:js_ph}  , we get a diffusion equation for $\mu^A_s$ which is 
\begin{align}
    \frac{\partial^2\mu^A_s}{\partial x^2} = \frac{\mu^A_s}{l_s^2}
    \label{eq:diffeq-mus},
\end{align}
and where $l_s=\sqrt{{\sigma\hbar}/{4e^2\Gamma}}$ is the characteristic length for spin flip relaxation.  \eqn{eq:jc}, \eqn{eq:js} and \eqn{eq:js_ph}  completely describe the charge and spin current flow inside the left metallic lead (see \Fig{fig:setup}). The spin current in the $\hat{\textbf{x}}$ direction leads to higher spin accumulation at interface A inside the metal. As a result,  polarized spins are injected into the magnetic insulator across the metal-insulator interface.

In the magnetic insulator, 
the injected spin current will be transported via magnons.  For system sizes larger than the magnon diffusion length, the generated magnon current will be diffusive in nature, much like the transport in the metal\cite{duineSpintronicsMagnonBoseEinstein2015}. Therefore, the equations governing spin transport in the magnetic insulator are similar to those of electrons in the metallic lead (see \eqn{eq:js} and \eqn{eq:js_ph}) and are given by \added{\cite{duineSpintronicsMagnonBoseEinstein2015}}
\begin{subequations}
\begin{align}
    j_s^x = -\frac{\sigma_s}{\hbar}\frac{\partial\mu_m}{\partial x}&- L_{SSE}\frac{\partial T_m}{\partial x},\label{eq:js_mag}\\
    \frac{\partial j_s^x}{\partial x}=-&\Gamma_m \mu_m,
\label{eq:js_mag_ph}
\end{align}
\end{subequations}
where  $\sigma_s$ is the spin conductivity of the insulator, $\mu_m$ is the non-equilibrium magnon chemical potential and $L_{SSE}$ is the bulk spin Seebeck coefficient. While the chemical potential for magnons is zero in equilibrium, upon injection of spins into the magnetic insulator, $\mu_m$ will develop a non-zero spatial profile similar to spin accumulation $\mu_s^A$ (\eqn{eq:diffeq-mus}). It turns out that having a non-trivial profile for $\mu_m$, with a finite spatial gradient, is crucial to bring BCD-induced magnon transport to first order in the temperature gradient.

The spatial profile for $\mu_m(x)$ can be arrived at by using \eqn{eq:js_mag} and \eqn{eq:js_mag_ph} to obtain the diffusion equation,
\begin{equation}
    \frac{\partial^2\mu_m}{\partial x^2} = \frac{\mu_m}{l_m^2},
    \label{eq:mu_m-diff}
\end{equation}
where $l_m=\sqrt{\sigma_s/\hbar \Gamma_m}$ describes the magnon diffusion length. 
In deriving \eqn{eq:mu_m-diff}, we have assumed a linear profile for the temperature $T_m(x)$ inside the magnetic insulator, such that $T_m(x) = T_0+\nabla T_m x$, where $\nabla T_m = (T_B-T_A)/L_x$, $T_0=(T_A+T_B)/2$, and ${\partial^2T_m}/{\partial x^2}=0$.

At the B end, this spin current can be converted into the charge current in metal, or equivalently an electric potential through the Inverse Spin Hall Effect (ISHE). Governing equations for this phenomenon will be 
\begin{subequations}\label{eq:jB_metal}
\begin{align}
     j_c^y  =& ~\;\frac{\sigma_{SH}}{2e} \frac{\partial \mu_s^B}{\partial x} \label{eq:jc_B}    \\
    j_s^x  = &-\frac{\sigma\hbar}{4e^2}\frac{\partial \mu_s^B}{\partial x}.
    \label{eq:js_B}
\end{align}
\end{subequations}
To account  for spin-flip relaxation, the above equations will also be supplemented by a diffusion equation for $\mu^B$:
\begin{align}
    \frac{\partial^2\mu^B_s}{\partial x^2} = \frac{\mu^B_s}{l_s^2}
    \label{eq:diffeq-mus_B},
\end{align} 
The boundary conditions that accompany \eqn{eq:diffeq-mus}, \eqn{eq:mu_m-diff} and  \eqn{eq:diffeq-mus_B} are  (i) currents in the $\hat{\textbf{x}}$ direction vanishes at the metal-vacuum interfaces perpendicular to  $\hat{\textbf{x}}$ (see \Fig{fig:setup}), and (ii) the currents at the metal-sample interface at the two ends (\Fig{fig:setup}) are dominated by the spin-exchange couplings between the metal and the sample and are given by\cite{duineSpintronicsMagnonBoseEinstein2015}
\begin{subequations}
    \begin{align}
        j_s^x(\sminus\tfrac{L_x}{2})=j_{sA}^{int}=\frac{\sigma_s^{int}}{\hbar\Lambda}\Big(\mu_s^A(\sminus\tfrac{L_x}{2})-\mu_m (\sminus\tfrac{L_x}{2})\Big)
    \label{eq:js_Ax}\\
        j_s^x(\tfrac{L_x}{2})=j_{sB}^{int}=\frac{\sigma_s^{int}}{\hbar\Lambda}\Big(\mu_m(\tfrac{L_x}{2})-\mu_s^B(\tfrac{L_x}{2})\Big),
    \label{eq:js_Bx}
    \end{align}
\end{subequations}
where $\sigma_s^{int}$, the interface spin conductivity 
\cite{duineSpintronicsMagnonBoseEinstein2015}.
Given these boundary conditions, the solution for the magnon chemical potential inside the magnetic sample is  (see, \app{ap_chem_pot})
\begin{equation}
\mu_m(x)=\frac{\mu_m(\minus\tfrac{L_x}{2}) \sinh \big(\tfrac{L_x-2x}{2l_m}\big)+\mu_m(\tfrac{L_x}{2}) \sinh \big(\tfrac{L_x+2x}{2l_m}\big)}{\sinh(L_x/l_m)};\label{eq:mu_m_x}
\end{equation}
a typical plot for $\mu_m(x)$ is shown in \Fig{fig:setup}.
The interface currents $j_{sA}^{int}$ and $j_{sB}^{int}$ can be tuned using the external electric field $\mathcal{E}$ used to set up the spin current in the metal (see \eqn{eq:j_metal}). Hence, the magnon chemical potential profile can be controlled by changing the applied electric field $\mathcal{E}$.  The exact dependence of the interface current on the applied electric field is derived in  \app{ap_chem_pot}.

In summary, \eqn{eq:mu_m_x} shows that the method of spin injection can generate a spatially varying chemical potential for magnons inside the magnetic insulator, which, as we will demonstrate, can be utilized to access BCD information in the linear order of temperature gradient.

 \section{Magnon Hall current induced by temperature gradient and spin-injection}\label{sec:BTE}
In this section, we {deploy} the setup described in the previous section to extract the Berry Curvature Dipole (BCD) in magnetically ordered systems.
 
A non-vanishing Berry Curvature ($\boldsymbol{\Omega}$) of magnon bands induces a transverse spin current $j^y_{\perp}$ in the $\hat{\textbf{y}}$ direction as a response to a longitudinal spin current $j_s^x$ in the $\hat{\textbf{x}}$ direction. The exact expression for the transverse spin current density is obtained in \cite{matsumotoRotationalMotionMagnons2011} in terms of a non-equilibrium, position-dependent magnon distribution function $\rho$ and is given by
\begin{equation}
j^y_\perp \!= -\frac{1}{V}\frac{\partial}{\partial x}\!\sum_{n,\,\textbf{k}}\Omega_n(\textbf{k})\!\!\int_0^\infty \!\!d\veps\,\rho(E_n(\textbf{k})-\mu_m(x)+\veps,\,T_m(x)) ,
\label{eq:jy_orig}
\end{equation}

 where $\boldsymbol{\Omega}_n (\boldsymbol{k})$ and $E_n (\textbf{k})$ are the Berry Curvature and the dispersion for the $n$-th band respectively.  The symbol $V$ denotes the area of the thin film. The sum in the index $n$ goes over bands having spin angular momentum pointing along the same direction, as a result the contribution from each band adds up with the same sign. When there are bands having spin angular momentum in the opposite direction, their contribution to the total spin current should be added to \eqn{eq:jy_orig} with a negative sign.
The additional integral over the energy-like parameter $\veps$ appears as a consequence of introducing a confining potential used to derive the expression for transverse current $j^y_\perp$ (\cite{matsumotoRotationalMotionMagnons2011}).

 We employ the Boltzmann Transport Equation (BTE)  to obtain the non-equilibrium distribution for magnons in the presence of temperature and chemical potential gradients. The BTE in the relaxation time approximation is given by
 \begin{equation}
 \frac{\partial\rho}{\partial t} + 
 \dot{\bl{x}}\cdot\frac{\partial \rho}{\partial \bl{x}} + 
 \dot{\bl{k}}\cdot \frac{\partial \rho}{\partial \bl{k}}= \frac{\rho_0(\veps)-\rho}{\tau}  \label{eq:bte}. 
 \end{equation}
 Here $\rho_0(\veps)\equiv n_B(E_n(\textbf{k})-\mu_m(x)+\veps, T_m(x))$, where $n_B(x,T) = (\exp(x/T)-1)^{-1}$ is the Bose-Einstein distribution defined locally and $\tau$ is the relaxation time. We look for a steady state solution in time to the above equation by setting $\partial\rho/\partial t =0$.  Hence the equation is reduced to the following form 
\begin{equation}
    \dot{\bl{x}}\cdot\frac{\partial \rho}{\partial \bl{x}}+ \dot{\bl{k}} \cdot\frac{\partial \rho}{\partial \bl{k}}=\frac{\rho_0(\veps)-\rho}{\tau}.
\label{eq:bte_std}
\end{equation}

An iterative solution \citep[p.~658]{kittel2018introduction} of the above equation up to $\mathcal{O}(\nabla^2)$ is given by \begin{equation}
\rho = \rho_0(\veps) -\tau\,  \dot{\bl{x}}\cdot\frac{\partial \rho_0(\veps)}{\partial \bl{x}}-{\tau}\,\dot{\bl{k}} \cdot\frac{\partial \rho_0(\veps)}{\partial \bl{k}}+\mathcal{O}(\nabla^2)\label{eq:rho}
\end{equation}
where the velocity $\dot{x}$ can be derived from band dispersion as $\dot{\bl{x}}=\hbar^{-1}{\partial E_n(\textbf{k})}/{\partial\k}$.
When time-reversal symmetry is present in the magnon sector, the Berry curvature is anti-symmetric under $\k\to-\k$, i.e., $\Omega_n(\textbf{k})=-\Omega_n(-\textbf{k})$. Additionally, the first and third terms in \eqn{eq:rho} are symmetric under $\k \to -\k$ while the second term is anti-symmetric.
Therefore, the integrals over the BZ coming from the contributions of the first and third term to \eqn{eq:jy_orig} will vanish and only the integral from the second term will contribute to Hall current $j_y$, which now becomes. \begin{align}
 \!\!\!    j^y_{\perp} =&- \frac{\tau}{\hbar V} \sum_{n,\textbf{k}}\Omega_n(\textbf{k})\;\frac{\partial E_{n}}{\partial k_x}\int_0^\infty d\veps\;\frac{\partial^2 \rho_0(\veps)}{\partial x^2}+ \mathcal{O}(\nabla^3).
\end{align}
We also note that the spatial variation in $\rho_0$ is only through $\mu$ and $T$, so we  write $\partial_x=(\nabla \mu)\;\partial_\mu+(\nabla T)\;\partial_T$. Therefore, the magnon Hall current \eqn{eq:jy_orig}, up to cubic order, is given by
\begin{align}
     j^y_{\perp} =&- \frac{\tau}{\hbar V} \sum_{n,\,\textbf{k}}\Omega_n(\textbf{k})\;\frac{\partial E_{n}}{\partial k_x}\int_0^\infty d\veps\;\Big[\nabla T^2 \frac{\partial^2 \rho_0(\veps)}{\partial T_0^2}\notag\\
     &~~~~~~~~~~~~+2\nabla T\nabla \mu \frac{\partial^2\rho_0(\veps)}{\partial T_0\partial\mu_0}+\nabla \mu^2 \frac{\partial^2 \rho_0(\veps)}{\partial \mu_0^2}\Big].\label{eq:jy_pert}
\end{align}
We evaluate the above integrals using two known functions associated to Bose-Einstein distribution function, $c_0(\rho_0)\equiv \!\partial/\partial\mu_0\!\int_0^\infty\! d\veps \rho_0(\veps)=\rho_0$, and $c_1(\rho_0)\equiv \partial/\partial{T_0}\!\int_0^\infty d\veps \rho_0(\veps)=(1+\rho_0)\ln(1+\rho_0)-\rho_0\ln\rho_0$,  where $\rho_0=n_B(E_n(\k)-\mu_0,T_0)$ (for derivation, see \cite{matsumotoRotationalMotionMagnons2011}) and  get
\begin{align}
    j^y_{\perp} \!=& \frac{\tau}{\hbar V} \!\sum_{n,\,\textbf{k}}\Omega_n\tfrac{\partial E_{n}}{\partial k_x}\Big[\nabla T^2 \tfrac{\partial c_1}{\partial T_0} +2\nabla T\nabla \mu \tfrac{\partial c_1}{\partial \mu_0}+\nabla \mu^2 \tfrac{\partial c_0}{\partial \mu_0}\Big].\label{eq:jy_pert_cs}
\end{align}
 We further simplify the above expression using functional relations, ${\partial c_{0,1}}/{\partial\mu_0}=-{\partial c_{0,1}}/{\partial E_n}$ and ${\partial c_{0,1}}/{\partial T_0}=-({E_n}/{T_0}){\partial c_{0,1}}/{\partial E_n}$ and replacing the sum over the first Brillouin Zone with an integral to arrive at
\begin{equation}
    j^y_{\perp} =- \frac{\tau}{\hbar V} {\sum\limits_{n}\!\!\int_{\text{\tiny BZ}}}\Omega_n\Big[ \nabla T^2 \tfrac{E_{n}}{T_0} \tfrac{\partial c_1}{\partial k_x}+2\nabla T\nabla \mu \tfrac{\partial c_1}{\partial k_x} +\nabla \mu^2 \tfrac{\partial c_0}{\partial k_x}\Big].\label{eq:jy_pert_cs_2nd}
\end{equation}
We find the total transverse magnon current get contributions from three different type of currents having distinct origins. Thus, we write the  final expression for magnon Hall current as
\begin{equation}
    j^y_{\perp}= j_{\text{\tiny SNE}} +j_{\text{\tiny bl}}+j_{\mu}.\label{eq:jy_final}
\end{equation}
In the above equation, the first term $j_{\text{\tiny SNE}}$ describes the spin-Nernst effect in topologically-trivial magnets and is nonlinear in temperature gradient. Ref. \cite{kondoNonlinearMagnonSpin2022} has 
shown $j_{\text{\tiny SNE}}$ to be proportional to the so-called Extended Berry Curvature Dipole (EBCD) of magnon bands, which can be seen by writing
\begin{equation}
\frac{j_{\text{\tiny SNE}}}{\nabla T^2}=\;\frac{\tau}{\hbar V}D^{(ext)}_{xy},
\end{equation}
where $\Dex$ denotes the EBCD, and is defined as
\begin{align}
    \Dex=&-  \sum\limits_{n}\!\!\int_{\text{\tiny BZ}}\Omega_n(\textbf{k}) \frac{E_{n}}{T_0} \frac{\partial c_1}{\partial k_x}\nonumber\\
    =&\;\frac{1}{T_0 }\sum_n  \int_{\text{\tiny BZ}} c_1(\rho_0)\frac{\partial (E_n({\textbf{k})}\Omega_n\textbf{(k))}}{\partial k_x}.\label{eq:extended_BCD}
\end{align}
To obtain the last expression for $\Dex$, we have shifted the differential operator $\partial_{k_x}$ from $c_1$ to $E\Omega$ using integration by parts along with periodic boundary conditions of the Brillouin Zone. 

{In contrast, the last two terms in \eqn{eq:jy_final} are new and arise only when a $\mu$-gradient is applied.} 
The third term in \eqn{eq:jy_final}, contributing to the Hall current, depends non-linearly on the $\mu$-gradient, and defines the ratio
\begin{equation}
\frac{j_{\mu}}{\nabla \mu^2}=\;\frac{\tau}{\hbar V}D^{(0)}_{xy},\label{eq:jmu_fin}
\end{equation}
where $\Do$ is the integral over BCD distribution in the x-direction and given by
\begin{equation}
\Do= \sum_n  \int_{\text{\tiny BZ}} c_0(\rho_0)\frac{\partial\Omega_n\textbf{(k)}}{\partial k_x}.\label{eq:Dxy0}
\end{equation}
{The above term provides a route to access the true BCD of magnon, even in the absence of a $T$-gradient.}

{The most interesting term in \eqn{eq:jy_final}  is the bilinear term $j_{\text{\tiny bl}}$ which depends linearly on both} $\nabla T$ and $\nabla\mu$ (see  \eqn{eq:jy_pert_cs_2nd}) and arises only when a $\mu$-gradient and $T$-gradient are applied simultaneously. Unlike the other two terms,  the magnon-current $j_{\text{\tiny bl}}$ changes direction  when the sign of either $\nabla T$ or $\nabla\mu$ is reversed. We name the current described by the $j_{\text{\tiny bl}}$-term the bilinear Hall current and  dub the phenomenon itself as the bilinear Hall response. Writing the bilinear response as 
\begin{equation}
    \frac{j_{\text{\tiny bl}}}{\nabla T\nabla\mu} = \frac{2\tau}{\hbar V}\Df,
    \label{eq:jmix_fin}
\end{equation}
we discover another moment of the BCD distribution
\begin{equation}
   \Df= \sum_n  \int_{\text{\tiny BZ}} c_1(\rho_0)\frac{\partial\Omega_n\textbf{(k)}}{\partial k_x}
    \label{eq:Dxy}
\end{equation}
oriented along the x-direction.
The contribution of $j_{\text{\tiny bl}}$ to the total transverse current $j^y_\perp$ shows that there is indeed a linear component in $j^y_\perp$ that will scale proportionately with $\nabla T$ when $\nabla \mu$ is held fixed in experiments, while the proportionality coefficient will be given by the BCD of magnon bands.

\eqn{eq:jmu_fin} and \eqn{eq:jmix_fin}  are two of the main results of our work.
It is interesting now to contrast {our} approach, culminating in equations \eqn{eq:Dxy0} and \eqn{eq:Dxy}, with recent results regarding the measurement of BCD for magnons via the Hall effect. 
In particular, Kondo et.al.\cite{kondoNonlinearMagnonSpin2022} considered only $T$-gradients and found that the extended BCD, a novel but alternate BCD-like quantity defined in \eqn{eq:extended_BCD}, determines the leading order contribution to transverse current.
In contrast, our approach of spin-injection assisted magnon-Hall effect is able to uncover the true BCD, i.e., the integrals of $\partial_{k_x}(\Omega_n(\k))$ instead of $\partial_{k_x}(E_n(\k)\Omega_n(\k))$. Furthermore, our expressions for the Berry curvature dipoles in \eqn{eq:Dxy0} and \eqn{eq:Dxy} are direct analogues of the electron BCD from \eqn{eq:BCD_electron}. \emph{To the best of our knowledge, our proposal so far is the only work in the context of magnons that captures the definition for BCD originally discovered for electrons} \cite{sodemannQuantumNonlinearHall2015}.

 Thus, we have calculated the magnon Hall current in the presence of simultaneous temperature ($T$) and magnon chemical potential ($\mu$) gradients and found two new contributions to the current which can be measured through experiments. Notably, one of these contributions scales linearly with $T$-gradient and probes the BCD in \eqn{eq:Dxy}, while the other scales non-linearly with $\mu$-gradient and probes the BCD in \eqn{eq:Dxy0}. In the subsequent sections, we will calculate these Berry curvature dipoles as well as the extended Berry curvature dipole \eqn{eq:extended_BCD} for various magnon systems.
 \section{Models}\label{sec:Models}
In this section, we study magnetically ordered phases for interacting spin Hamiltonian defined on three types of 2D lattice systems, namely the honeycomb lattice, the kagome lattice and the dice lattice under the approximations of linear spin wave theory (LSWT)\cite{kittel1963quantum,stancil2009spin,toth2015linear}.
Specifically, we study the following five cases -- the antiferromagnetic and ferromagnetic phases on the honeycomb lattice in subsection \ref{Sec:HC}, the ferromagnetic phase on the kagome lattice in subsection \ref{Sec:kagome}, and the ferrimagnetic and ferromagnetic phases on the dice lattice in subsection \ref{Sec:dice}.  We choose the LSWT route over more sophisticated approaches to study the above phases, since it provides a conceptually transparent (yet powerful) way to identify magnonic excitations in magnetically ordered spin systems and has been widely applied to explain magnonic phenomena in several models with great success\cite{toth2015linear}.

The LSWT approximation proceeds by implementing the
 Holstein-Primakoff transformation of the spin operators in the Hamiltonian and retaining terms up to linear order in spin length $S$ while dropping terms that are $\mathcal{O}(1/S)$. Next, using the linear Hamiltonian in $S$, we calculate the Bloch Hamiltonian $\mathcal{H}(\textbf{k})$, for each model, and identify the magnon bands by computing the eigen-spectrum of $\mathcal{H}(\textbf{k})$, while the Berry Curvature $\Omega_n(\k)$ for the $n$-th band {is obtained from the eigenvectors of} $\mathcal{H}(\textbf{k})$. With the information of magnon bands and Berry curvature, we are readily able to calculate the  BCDs -- $D^{(0)}_{xy}$, $D^{(1)}_{xy}$  and  the EBCD, $D^{(ext)}_{xy}$ for each spin model, as a function of several Hamiltonian parameters, such as nearest neighbour exchange interaction strengths, spin anisotropy, as well as temperature.

While exploring the space of these parameters, we deliberately stay in regimes where the magnon phases are not topological, since our goal is to access purely geometric effects in transport properties. 
Doing so will ensure that currents originating from geometry, such as the bilinear Hall current (\eqn{eq:jy_final}), are the dominant contributors to the total transverse current in magnon transport experiments.
\subsection{Honeycomb Lattice}\label{Sec:HC}
\begin{figure}[t]
\includegraphics[scale=1.04]{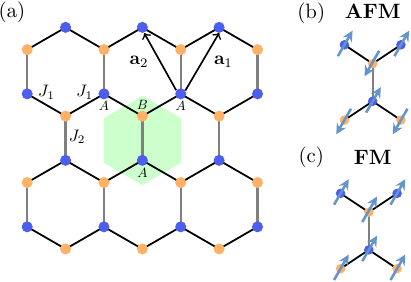}
    \caption{{Honeycomb lattice}: (a) The unit cell having two sites $A$ and $B$ is shown by shaded hexagon and the primitive lattice vectors are represented by
$\textbf{a}_1$ and $\textbf{a}_2$. (b) and (c) shows the orientation of spins on sub-lattices $A$ and $B$ for 
anti-ferro and ferro-magnetic order, respectively.   The   coupling strength for slanted (vertical) bonds is  $J_1 (J_2)$.
}\label{fig:honeycomb}
\end{figure}

We begin with the formalism for finding the BCD of magnonic excitations for the spin Hamiltonian
\begin{align}
\begin{split}
    H = & J_1 \sum_{\langle i,j\rangle\in \text{ slant}}\textbf{S}_i\cdot \textbf{S}_j +J_2 \sum_{\langle i,j\rangle \in \text{vertical}}\textbf{S}_i\cdot\textbf{S}_j \\     
    &~~~~~~ ~~~~~~-\kappa_A \sum_{i\in A} (\textbf{S}_i^{z})^2-\kappa_B\sum_{j\in B} (\textbf{S}_j^{z})^2
\end{split} \label{eq:HC_Ham}
\end{align}
defined on the honeycomb lattice (see \fig{fig:honeycomb}); the ground state for which is an antiferromagnet (AFM) when $J_{1,2}>0$ and ferromagnet for $J_{1,2}<0$.  
The formalism we develop in this section is quite general and will be applied to calculate the magnon BCD for the spin models defined on the other lattices as well.  

In the Hamiltonian (\eqn{eq:HC_Ham}), the first (second) term, with coupling strength $J_1 (J_2)$, represents a Heisenberg-type interaction between nearest neighbour spins along the slanted (vertical) bonds (see \fig{fig:honeycomb}). The third and fourth terms introduce an easy axis anisotropy of strength $\kappa_{A},\kappa_{B} > 0$, along the out-of-plane direction $\boldsymbol{\hat{z}}$, for spins on sublattices $A$ and $B$ respectively. The anisotropy strength is usually much lower than $J_{1,2}$, i.e. $\kappa_{A/B}\ll J_{1,2}$, but helps in stabilizing magnetic order in 2D-systems\cite{aharoni2000introduction}. When $J_1=J_2$, the honeycomb lattice is considered to be un-distorted having a threefold-rotational symmetry ($\mathrm{C}_3$). Experimentally, we can introduce distortions in the honeycomb lattice by applying strain which we can model theoretically by tuning $J_2$ away from $J_1$. Doing so will allow us to predict the effect of strain on the BCDs in the next section.
\subsubsection{Honeycomb Antiferromagnet}\label{Sec:model_HC_AF}
The AFM order for the Hamiltonian (\eqn{eq:HC_Ham}) can be obtained by setting $J_{1,2}>0$ and $\kappa_A=\kappa_B$.
The AFM ground state has spin configurations on the $A$ and $B$ sub-lattice sites pointing along opposite directions that are parallel and anti-parallel to the $\bf{z}$-axis, or vice-versa. We show one such AFM configuration in \fig{fig:honeycomb} inset, and denote the magnitude of spin moment per site for the AFM configuration with $S$.
To access the magnon sector,
we first transform the Hamiltonian (\eqn{eq:HC_Ham}) from a spin-operator form into a bosonic operator form using the  Holstein-Primakoff (HP) transformations 
\begin{equation}\label{eq:hp_transform_afm}
\begin{split}
 & \,\textbf{S}^z_i = S - a^\dagger_i a_i \\
 &  \textbf{S}^+_i = (2S - a^\dagger_i a_i)^{1/2}\,a_i \\
 &  \textbf{S}^-_i = a^\dagger_i \,(2S - a^\dagger_i a_i)^{1/2}
\end{split}
~~\Bigg |~~
\begin{split}
 & \textbf{S}^z_j\, = -S + b^\dagger_j b_j \\
 &  \textbf{S}^+_j = b^\dagger_j (2S - b^\dagger_j b_j)^{1/2} \\
 &  \textbf{S}^-_j =  ( 2S - b^\dagger_j b_j)^{1/2}\,b_j
\end{split}
\end{equation}
where $a_i (b_j)$ and $a_i^\dagger (b_j^\dagger)$ are respectively the magnon annihilation and creation operators for the $i$ ($j$)-th site  corresponding to the $A$ ($B$) sub-lattice. The operators obey the bosonic commutation relations $[a_i,b_j]=[a^\dagger_i,b^\dagger_j]=0$, $[a_i,a^\dagger_j]=\delta_{ij}=[b_i,b^\dagger_j]$ and consequently preserve the required spin commutators $[\textbf{S}^\alpha,\textbf{S}^\beta]=i\hbar\,\epsilon^{\alpha\beta\gamma}\,\textbf{S}^{\gamma};\alpha,\beta,\gamma\in {x,y,z}$. Along with the above transformations, we Fourier transform the Hamiltonian $H$ and keep terms only up to linear order in  $S$ to get the following Bloch Hamiltonian describing non-interacting magnons  
\begin{equation}
H = \frac{1}{2} \sum_{\k \,\in \textrm{BZ}} \Psi^\dagger(\textbf{k})\mathcal{H}_{\k} \Psi(\textbf{k}).\label{eq:Mag_Ham_HCAF}
\end{equation}
Here, $\k$ represents Bloch momenta, and $\Psi^\dagger(\textbf{k})=(a^\dagger_{\k}\;b^\dagger_{\k}\;a_{-\k}\;b_{-\k})$; the momentum-space operators $a_\k$, $b_\k$ are related to their lattice versions $a_i$, $b_i$ via the Fourier transformations
\begin{equation}
   \!\! a_{\textbf{k}} = \sum_{\textbf{R}_i}a_i \, e^{-i\textbf{k}\cdot(\textbf{R}_i + \textbf{r}_{A})},~~    b_{\textbf{k}} = \sum_{\textbf{R}_i}b_i \, e^{-i\textbf{k}\cdot(\textbf{R}_i + \textbf{r}_{B})}\label{eq:Fourierab}
\end{equation}
where $\textbf{R}_i$ is the lattice position vector of an unit cell  and $\textbf{r}_{A}$ ($\textbf{r}_{B}$) is the position of  $A$ ($B$) sub-lattice site inside the unit cell relative to $\textbf{R}_i$. The $4\times 4$ coefficient matrix $\mcH_\k$ has the form
\begin{equation}
    \mathcal{H}_\textbf{k}=\begin{pmatrix}
                            d & 0 & 0 & \gamma_\textbf{k}\\
                            0 & d & \gamma^*_\textbf{k} & 0\\
                            0 & \gamma_\textbf{k} & d & 0\\
                            \gamma^*_\textbf{k} & 0 & 0 & d 
                            \end{pmatrix}\label{eq:hc_afm_bdg}
\end{equation}
where $d = (2J_1  +  J_2  + 2\kappa )$ and the off diagonal coupling $\gamma_{\k} = 2J_1\cos(k_x/2)e^{ik_y/(2\sqrt{3})} +J_2 e^{-ik_y/\sqrt{3}}$.
The coupling $\gamma_{\k}\neq 0$ introduces boson non-conserving terms, such as $ab$ and $a^\dagger b^\dagger$, in $H$ making it a  Bogoliubov–de Gennes (BdG) Hamiltonian \cite{kondoNonHermiticityTopologicalInvariants2020,matsumoto2014thermal}.
Therefore, {$\Hk$} cannot be diagonalized via unitary transformations of $\Psi(\k)$, since such a transformation will not preserve the bosonic commutation relations obeyed by $a^\dagger$, $a$, and $b^\dagger$, $b$. 
Hence to obtain the eigenmodes of $H$, one has to use \emph{paraunitary} matrices\cite{colpa1978diagonalization}, say $P$, to transform $\Psi(\k)$ so that the eigenmodes remain bosonic after the transformation. This entire analytical process can be simplified down to diagonalizing the matrix $\Sigma_z \mathcal{H}_\textbf{k}$ (where $\Sigma_z=\sigma_z\otimes\mathbb{I}_2$),  instead of $\mathcal{H}_\textbf{k}$\cite{kondoNonHermiticityTopologicalInvariants2020}.  The steps are carried out in \app{ap_spin_wv_theo}, and the resulting four BdG bands will have eigenvalues \begin{equation}
P^\dagger  \mathcal{H}_\textbf{k} P =\text{diag}(E_1({\k}),~E_2({\k}),~E_1(-{\k}),~E_2(-\k)) \label{eq:hc_afm_bdg_diag}\end{equation}
where $E_{1,2}(\k)=\sqrt{d^2-|\gamma(\k)|^2}$. We find the first two eigenvalues correspond to physical magnon bands which are also degenerate in this case. The remaining two eigenvalues are copies of the first two and arise as an artefact of diagonalizing BdG Hamiltonians. 
Therefore, it is sufficient for us to consider only the first two degenerate magnon modes, with dispersions $E_{1,2}(k)$, in the remaining calculations. The energy degeneracy of these two modes is a consequence of the underlying sub-lattice symmetry, which leaves the Hamiltonian (\eqn{eq:HC_Ham}) invariant under the interchange of $A$, $B$ sub-lattice labels when $\kappa_A = \kappa_B$.
Although, the magnon modes are degenerate they can still be distinguished since they carry opposite spin angular momentum in the z-direction. This is also evident from the structure of $\mcH(\k)$ as it decouples into two blocks when represented in the sub-lattice space . The decoupling allows us to isolate individual magnon modes despite their energy degeneracy and to calculate their BC distributions using standard approaches developed for non-degenerate bands (see \app{ap_spin_wv_theo}).

We find the Berry curvatures of these two magnon bands are {$\Omega(\k),-\Omega(\k)$}, where \begin{equation}
\!\!\!\Omega(\k)=\frac{dJ_1\,\big[J_1\sin(k_x)-2J_2 \sin(\frac{k_x}{2})\cos\big(\frac{\sqrt{3}k_y}{2})\big]}{4\sqrt{3}(d^2-|\gamma_{\k}|^2)^{3/2}}\label{eq:HC_AF_BC}
\end{equation}
and satisfies the time-reversal property $\Omega(\k)=-\Omega(-\k)$ implying the Chern numbers for both bands are zero. Since the two BdG magnon bands have equal and opposite spin angular momentum (see \app{ap_spin_wv_theo}),  the total magnon spin current in the transverse direction is {given by} the difference between the currents
\begin{align}
j^y_\perp = j^y_{\perp,1}-j^y_{\perp,2},
\end{align}
 where $j^y_{\perp,1}$ ($j^y_{\perp,2}$) are contributions from the 1st (2nd) magnon band given by the expression in \eqn{eq:jy_orig}. Therefore, the two BCDs and EBCD for this system is determined to be
\begin{align}
\begin{split}
D^{(0,1)}_{xy}=& \;2\int_{\text{\tiny BZ}} c_{0,1}(\rho_0) \,\frac{\partial\Omega}{\partial k_x}, \\
\Dex=& \;2\int_{\text{\tiny BZ}} c_{1}(\rho_0)\, \frac{\partial (E\Omega)}{\partial k_x} 
\end{split}\label{eq:HC_dipoles}
\end{align}
which we use to numerically calculate BCDs and EBCD for the honeycomb anti-ferromagnet in the results section.

\subsubsection{Honeycomb Ferromagnet}\label{Sec:model_HC_F}
The ferromagnetic (FM) state on the honeycomb lattice can be realized by setting $J_{1,2}<0$ in the Hamiltonian \eqn{eq:HC_Ham}.
We also choose different values for easy axis anisotropies {$\kappa_A$, $\kappa_B$} for the two sub-lattices $A$, $B$. Doing so, breaks the sub-lattice interchange symmetry, $A\leftrightarrow B$, in the Hamiltonian, leading to magnon bands that are energetically gapped everywhere in the BZ. Having gapped bands, allows us to easily calculate the Berry curvature $\Omega_n(\k)$ for each isolated band without having to handle the computation of BC at degenerate points in the BZ. Under the Holstein-Primakoff transformations for ferromagnets\cite{kittel1963quantum}, spins on both sub-lattices $A$ and $B$ transform similarly as follows
\begin{equation}\label{eq:HF_ferro}
\begin{split}
 & \,\textbf{S}^z_i = S - a^\dagger_i a_i \\
 &  \textbf{S}^+_i = (2S - a^\dagger_i a_i)^{1/2}\,a_i \\
 &  \textbf{S}^-_i = a^\dagger_i \,(2S - a^\dagger_i a_i)^{1/2}
\end{split}
~~\Bigg |~~
\begin{split}
 & \textbf{S}^z_j\, = S - b^\dagger_j b_j \\
 &  \textbf{S}^+_j = ( 2S - b^\dagger_j b_j)^{1/2}\,b_j \\
 &  \textbf{S}^-_j =  b^\dagger_j (2S - b^\dagger_j b_j)^{1/2}
\end{split}
\end{equation}
The symbols $a$, $a^\dagger$ and $b$, $b^\dagger$ etc. have the same meanings as discussed near \eqn{eq:hp_transform_afm}.
Transforming the spin operators in the Hamiltonian (\eqn{eq:HC_Ham}) using the above expressions and moving to the Bloch-momentum representation, we get the  magnon Hamiltonian for the ferromagnet: 
\begin{equation}\label{4.B.2}
    H = \sum_{\k\in \mathrm{BZ}}\Phi^\dagger(\textbf{k}) \mathcal{H}_{\k}\Phi(\textbf{k})
\end{equation}
where 
\begin{equation}\label{4.B.3}
    \mathcal{H}_{\k}= \begin{pmatrix}
        ~~d_A & \sminus\,\gamma_{\k}\\
        \sminus\,\gamma^*_{\k} & ~~d_B
    \end{pmatrix}
\end{equation}
with $\Phi^\dagger(\k)=(a^\dagger_\textbf{k},b^\dagger_\textbf{k})$, $d_{A/B}=2 J_1+J_2+2\kappa_{A/B}$, and  $\gamma_{\k}= 2J_1\cos(k_x/2)e^{ik_y/(2\sqrt{3})} +J_2 e^{-ik_y/\sqrt{3}}$. Unlike the antiferromagnetic case (\eqn{eq:hc_afm_bdg}), the Hamiltonian above does not contain any number non-conserving terms. Therefore, the band dispersions and wavefunctions for magnons can be found by directly diagonalizing the coefficient matrix $\mathcal{H}_\textbf{k}$. The resulting two magnon modes have band energies $E_{1,2}(\k)=(d_A+d_B)\mp [(d_A-d_B)^2+|\gamma_{\k}|^2]^{1/2} $ and corresponding Berry curvatures $\pm\Omega(\k)$ respectively, where
\begin{equation}
\Omega(\k)=\frac{(\kappa_A \sminus\,\kappa_B)J_1\,[J_1\sin(k_x)\sminus\,2J_2 \sin(\frac{k_x}{2})\cos\big(\frac{\sqrt{3}k_y}{2}\big)]}{4\sqrt{3}[(\kappa_A\sminus\,\kappa_B)^2+|\gamma_{\k}|^2]^{3/2}}.\label{eq:HC_F_BC}
\end{equation}
Since $\Omega(\k)$ is proportional to $(\kappa_A-\kappa_B)$, we need to set $\kappa_A\neq \kappa_B$ to get a non-zero Berry curvature distribution over the bands. Therefore, in order to set $\kappa_A\neq \kappa_B$ we need to break the sublattice symmetry present in the model. This is in contrast with the antiferromagnetic case where we found a finite BC without breaking the sublattice symmetry.
Having obtained $\Omega (\k)$ and $E_{1,2}(\k)$ above, we can use the same expressions derived in \eqn{eq:HC_dipoles} to compute the BCDs --$\Do$, $\Df$  and  the EBCD, $\Dex$ for the ferromagnetic state as well.

Due to the dependence of $\Omega$ on $J_{1,2}$ for antiferromagnet in \eqn{eq:HC_AF_BC} and ferromagnet in \eqn{eq:HC_F_BC}, we conclude that the magnon Hall current in both these phases can be strain-tuned, with the strain being modelled by the difference in the value of $J_1$ and $J_2$. In addition, we also find that this current is dependent on the easy axis anisotropy value $\kappa_{A,B}$. We {utilize these feature, to}  show in the results section that the magnon Hall response, and hence magnon BCDs and EBCD for the HC lattice, can be controlled {efficiently} using strain. 

\subsection{Kagome Lattice}\label{Sec:kagome}
\begin{figure}[t]
     \centering
\includegraphics[scale=0.9]{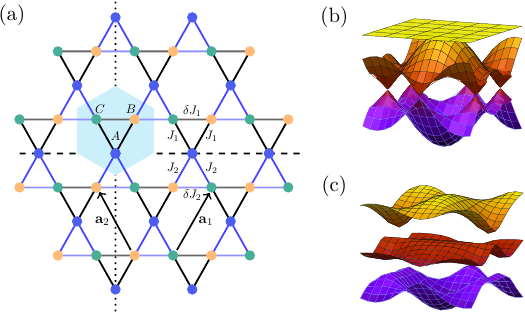}
    \caption{Kagome lattice: (a) A unit cell, containing three spins on sublattice $A$, $B$, and $C$, is represented by a shaded regular hexagon, with two lattice vectors $\textbf{a}_1$ and $\textbf{a}_2$  shown. The coupling strengths for slanted  and horizontal bonds belonging to down-triangle (up-triangle) are $J_1$($J_2$) and $\delta J_1$ ($\delta J_2$) respectively. The relative positions of nearest neighbour spins within the unit cell are $\textbf{r}_{AB}= \hat{x}/4 +\sqrt{3}\,\hat{y}/4$, $\textbf{r}_{AC}= -\hat{x}/4 +\sqrt{3}\,\hat{y}/4$ and $\textbf{r}_{CB}=\hat{x}/2 $. (b) The magnon band structure for the pristine kagome ferromagnet ($\delta=1$, $J_1=J_2$). (c) Representative band structure for the kagome ferromagnet under deformation when $\delta=0.4$, $J_1=0.4$, $J_2=0.6$. The dashed and dotted lines represent  the vertical and horizontal lines of reflection symmetry respectively. }
    \label{fig:kagome_ferro}
\end{figure}

The next system we consider for investigating the BCD contributions to transverse magnon current is a model on the kagome lattice with a ferromagnetic ground state. The kagome lattice, as illustrated in \Fig{fig:kagome_ferro}, is a hexagonal Bravais lattice with corner-sharing triangles. Its unit cell is a regular hexagon containing three distinct sites denoted as $A$, $B$, and $C$ corresponding to the vertices of the triangles.
Exploring spin phases on the kagome lattice is one of the most active areas of research in the spin-physics community due to the myriad of exotic phases that the lattice can support.
In particular, kagome lattice ferromagnets are known to host excitations with topological band structure whose Hall effect has been detected in {recent} experiments \cite{topo_mag_mat,topo_mag_room_temp}. Several compounds, such as  Cu$_9$X$_2$(cpa)$_6$ (X $=$ F, Cl, Br; cpa $=$ 2-carboxypentonic acid anion) \cite{triangle_kagome_magnons}, have been identified that have a 2d kagome lattice {geometry} whose couplings can be tuned experimentally. The above makes the kagome lattice ferromagnet a potent system in which the geometrical spin Hall current can be experimentally observed.  We consider the following Hamiltonian to model the kagome ferromagnet
\begin{equation}\label{k1.1}
\begin{split}
H = H_{\wedge} + H_{\vee} + H_{-}+ H_{an}
\end{split}
\end{equation}
where 
\begin{equation}\label{1.2}
\begin{split}
    & H_{\wedge} = -J_1 \sum_{\nn{i}{j} }^\wedge\textbf{S}_i\cdot\textbf{S}_j,~~~~ H_{\vee} = -J_2 \sum_{\langle i,j\rangle}^\vee\textbf{S}_i\cdot\textbf{S}_j \\
    &H_{-} = -\delta\Bigg(J_1 \sum_{\langle i,j\rangle}^-\textbf{S}_i\cdot\textbf{S}_j+J_2 \sum_{\nn{i}{j}}^-\textbf{S}_i\cdot\textbf{S}_j\Bigg)\\
    & H_{an} = -\kappa \sum_i (\textbf{S}_i^z)^2. 
\end{split}
\end{equation}
The term $H_\wedge$ ($H_\vee$) denotes the spin-spin interactions for the \emph{slanted} bonds of the up (down)-pointing triangles in \fig{fig:kagome_ferro} having strength $J_1$ ($J_2$). The $H_{-}$ term introduces strain in the model by modulating the strengths $J_1$ and $J_2$ for the horizontal bonds belonging to the up and down pointing triangles, respectively, using the parameter $\delta$. The value of $\delta=1$ corresponds to the kagome lattice with no strain. In this limit, the kagome lattice has a threefold rotational symmetry $\mathrm{C}_3$ about an axis through the centroid of the unit cell.
We vary $\delta$ away from \emph{one} to demonstrate how strain along any given direction can generate a non-vanishing BCD in this model. For a fixed value of $\delta$, the extreme limits $J_1\gg J_2$ and $J_1\ll J_2$ correspond to isolated triangles of opposite orientations.  When $J_1=J_2$, the kagome lattice has an additional symmetry under reflection about a horizontal axis connecting the $A$ sites (see \fig{fig:kagome_ferro}) irrespective of $\delta$. For both $\delta = 1$ and $J_1=J2$, the lattice has $\mathrm{C}_3$ rotation and reflection symmetry about the same horizontal axis as well as their group products. The anisotropy term $H_{an}$ is introduced to stabilize, the magnetic order. Since we set the same easy axis anisotropies for all three sublattices to be same, the lattice has an additional reflection symmetry about the vertical axis (see dotted line in \fig{fig:kagome_ferro}) for all values of $\delta$ and $J_{1,2}$.

Analogous to the previous models, we apply LSWT to obtain the magnon Hamiltonian for this system as well.  For which, we perform the Holstein-Primakoff transformations followed by a Fourier transform for the three sites $A$, $B$, $C$ similar to \eqn{eq:HF_ferro}  and \eqn{eq:Fourierab} respectively. Here we state the HF and Fourier transformations only for the additional third site $C$ in the unit cell, which are
\begin{align*}
\begin{split}
\textbf{S}^z_l=&S\sminus\,c^\dagger_l c_l,\;
\textbf{S}^+_l \!=\!( 2S \sminus\, c^\dagger_l c_l)^{\frac{1}{2}}\,c_l,\;\textbf{S}^-_l \!=c^\dagger_l (2S \sminus\, c^\dagger_l c_l)^{\frac{1}{2}}
\end{split}\end{align*}
and
\begin{align}
c_{\,\textbf{k}} =& \sum_{\textbf{R}_l}c_l e^{-i\textbf{k}.(\textbf{R}_l + \textbf{r}_{C})},\label{eq:Fourierc}
\end{align}
respectively. Where $l\in C$ and various symbols $c$, $c^\dagger$, $\textbf{r}_C$  has similar meanings as discussed near \eqn{eq:hp_transform_afm}. Upon completion of this procedure, we get the following {Bloch} Hamiltonian
\begin{align}
H=\sum_{\k}\;\Phi^\dagger(\textbf{k})\mathcal{H_\textbf{k}}\Phi({\textbf{k}})\label{eq:kagome_magnon_ham}
\end{align}
where $
    \Phi^\dagger(\textbf{k})=\begin{pmatrix}
        a^\dagger_{\k} & b^\dagger_{\k} & c^\dagger_{\k}
    \end{pmatrix} $, the coefficient Hamiltonian 
\begin{equation}\label{eqn:Hk_kag}
    \mathcal{H}_\textbf{k} = \begin{pmatrix}
                   ~\; d_1 & \sminus\,\Delta_{1\k} & \sminus\,\Delta_{2\k} \\
                   \sminus\,\Delta_{1\k}^* & ~d_2 & \sminus\,\Delta_{3\k}\\
                   \sminus\,\Delta_{2\k}^* & \sminus\,\Delta_{3\k}^* & ~\;d_3
        
                    \end{pmatrix},
\end{equation}
$ d_1 = \; 2(J_1+J_2)+2\kappa$, $
d_2 = \; d_3 =(1+\delta) (J_1+J_2)+2\kappa$ and
\begin{equation}\label{k1.4}
\begin{split}
\Delta_{1\k} =&\; e^{-\frac{i k_x}{4}} (J_1 e^{\frac{i\sqrt{3}}{4}k_y} +J_2 e^{-\frac{i\sqrt{3}}{4}k_y}) \\
\Delta_{2\k} = &\;e^{\frac{i k_x}{4}} (J_1 e^{\frac{i\sqrt{3}}{4}k_y} +J_2 e^{-\frac{i\sqrt{3}}{4}k_y}) \\
\Delta_{3\k} = &\; \delta\big(J_1 e^{ik_x/2} +J_2 e^{-ik_x/2}\big).
\end{split}
\end{equation}
Similar to the ferromagnetic magnons on the honeycomb lattice, the magnon dispersion and Bloch magnons of the ferromagnetic-kagome lattice are obtained by diagonalizing the coefficient matrix $\mcH_\k$ in \eqn{eqn:Hk_kag} via unitary transformations. Since $\mcH_\k$ is a $3\times 3$ matrix, it is difficult to obtain analytical expressions for the eigenvectors of $\mcH_\k$ and Berry curvatures of the resulting three bands for arbitrary values of the coupling parameters $J_1$, $J_2$ etc.  Therefore, we use numerical diagonalization to obtain the spectrum of $\mcH_\k$ and the Wilson loop method, prescribed in \cite{FukuiChern2005, wang2020universal}, for calculating Berry curvature distributions over the bands. 
A typical magnon band structure for the undistorted kagome lattice (for $\delta =1$ and $J_1=J_2$) is presented in \fig{fig:kagome_ferro}(inset) showing three bands. The uppermost band is flat and touches the middle band at $\k=\textbf{0}$; while the lower two bands touch at $\pm\k_0$, where $\k_0 = (4\pi/3) \hat{x}$ for $\delta=1$, and moves along $x$-axis in the BZ  upon changing the value of $\delta$. The latter degeneracy of the lower two bands is due to the reflection symmetry about the horizontal axis shown in \fig{fig:kagome_ferro} and is removed when $J_1\neq J_2$ for any value of $\delta$. The degeneracy between the upper two bands is protected by $\mathrm{C}_3$ rotational symmetry and time reversal symmetry .
When $\delta$ is decreased from \emph{one}, the degeneracy at $k=0$ first splits into two band touchings at certain momenta $\pm k_*$, lying on the $y$-axis, and then lifts when $\delta$ becomes less than a critical value set by $J_1/J_2$.

We study the behavior of the BCDs of kagome ferromagnet under strain by tuning $\delta$ across \emph{unity}, {as well as, different values of the ratio $J_1/J_2$} and present the results in \Sec{sec:Results}.
\subsection{The Dice Lattice}\label{Sec:dice}
\begin{figure}[t]
\centering
\includegraphics[scale=1.0]{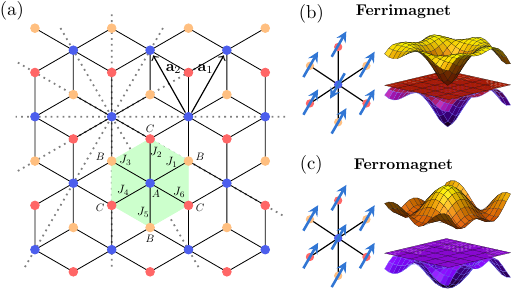}
\caption{The dice lattice: (a) A hexagonal unit cell containing three spins, one from each sub-lattice $A$, $B$ and $C$ is shown by a shaded region and the two primitive lattice vector are represented by $\textbf{a}_1$ and $\textbf{a}_2$. The centre spin of the unit cell, $A$ interacts with six spins, three from each sub-lattice $B$ and $C$ with coupling strength $J_{1-6}$. (b) and (c) show the spin orientations and the magnon band structure for pristine dice lattice (all coupling strengths equal) with ferri and ferro-magnetic orders, respectively. The dotted lines represent the lines of reflection symmetries of the pristine dice lattice.}\label{fig:Dice_lattice}
\end{figure}

Following the study of BCD for anti-ferromagnetic and ferromagnetic magnons,   
we shall now, in this section,  explore BCD of magnetic excitations of ferri-magnetic order. In ferrimagnetic order, nearest neighbours spins are oriented in opposite directions, yet the total spin (hence magnetization) of an unit cell is \emph{non-zero}. In this sense, ferrimagnetic order can be visualised as an intermediate between pure ferromagnetic and pure anti-ferromagnetic orders.
A fitting choice for studying ferri-magnetic order would be a lattice  commonly referred to as the Dice lattice or ${\cal T}^3$ lattice \cite{vidal1998aharonov} which has three sites in its unit cell (see \Fig{fig:Dice_lattice}). The dice lattice, also known to be the dual of kagome lattice \cite{sen2009frustrated}, is essentially a honeycomb lattice with an extra atom positioned at the centre of each hexagonal unit cell. Recently, the dice lattice has gained huge interest in tight-binding models for electron systems.  Due to the presence of a non-trivial flat band, dice lattice, shows a rich phase diagram as compared to graphene lattice and exhibits various  exotic phenomena like Klein tunnelling \cite{illes2017klein},  unconventional Anderson localization \cite{vidal1998aharonov},   unusual
magnetic-optical effect \cite{beugeling2014nontrivial,chen2019enhanced}, Majorana corner states and higher-order topology\cite{mohanta2023majorana}, mass-less Dirac-Weyl fermions with pseudospin $S=1$ \cite{bercioux2009massless}  and many more \cite{liu2023thermopower}.  Magnon models also have been studied on this lattice in various contexts such as geometric fluctuations\cite{jagannathan2012geometric},  collinear and non-collinear phases \cite{sahoo2018classical}. In addition to the above, we choose to study the dice lattice for three reasons. First, the unit cell of this lattice contains three distinct spins associated with sub-lattice sites $A$, $B$ and $C$, which will be required to stabilise the ferrimagnetic order.  
Second, this is the next simplest lattice to honeycomb in which both anti-ferromagnetic and ferromagnetic interactions between nearest neighbours can be studied without leading to frustrations in the ground state. For instance, introducing anti-ferromagnetic interactions allows us to access the ferrimagnetic order on the Dice lattice.
Finally, the dice lattice serves as a parent lattice which can be deformed (possibly using strain) to realize other lattices including the Honeycomb lattice (see \Sec{Sec:Merged_HC_model}), the coupled dumble chain lattice (see \Sec{Sec:dumble_model}) and the boat lattice ( \Sec{Sec:Boat_lattice_model}). Thereby, allowing us to study the BCD of magnons on multiple lattices in one shot. To access the ferri and ferromagnetic ground states on the dice lattice, we consider the following spin Hamiltonian.
\begin{align}
\begin{aligned}
H= &\sum_{\substack{i\in A\\\eta \in \{1\cdots6\}}} J_\eta \;\textbf{S}_{i}\cdot \textbf{S}_{i+\eta}-\kappa \sum_{i\in \text{sites}}  \;(\textbf{S}^{z}_{i})^2\\
\end{aligned}\label{eq:Ham_dice}
\end{align}
In the above Hamiltonian, the first term (summing over all $i \in A$ sites) describes spin-spin interactions along bonds (indexed by $\eta$) connecting $A$ sites to their nearest neighbor spins located on either $B$ or $C$ sites. The index $\eta\in\{1,\cdots\!,6\}$ represents directions pointing towards the six nearest neighbours of an $A$-type site, as shown in \fig{fig:Dice_lattice}, and enumerates the six bonds along those directions having interactions strengths $J_1,\cdots,J_6$, respectively. In the second term, we have introduced an easy axis anisotropy $\kappa$,  at all sites, for the stability of the magnetic order. We note that there is no interaction between $B$ and $C$ site spins in this Hamiltonian. The presence of an  anti-ferromagnetic interaction connecting $B$ and $C$ would effectively make it a triangular lattice and would lead to a frustrated ground state. 
\subsubsection{Dice Ferrimagnet}\label{Sec:dice_FE}
The ferrimagnetic order on this lattice (see \fig{fig:Dice_lattice}(b)) can be realised by setting all the couplings $J_{1-6}<0$  in \eqn{eq:Ham_dice}. 
As discussed earlier,  anti-ferromagnetic exchange interactions are required to establish ferrimagnetic order in the system. Thus, we use the Holstein-Primakoff transformations 
\begin{align}
\begin{split}
\textbf{S}^z_i\!=&\,S \sminus\, a^\dagger_ia_i,\;\;
\textbf{S}^+_i \!=\!( 2S \sminus\,a^\dagger_i a_i)^{\frac{1}{2}}a_i,\,~\textbf{S}^{\sminus}_a \!=a^\dagger_i (2S \sminus\, a^\dagger_i a_i)^{\frac{1}{2}}, \\
\textbf{S}^z_j\!=&\sminus\,S+b^\dagger_j b_j,\,
\textbf{S}^+_j \!=\!b^\dagger_j (2S \sminus\, b^\dagger_j b_j)^{\frac{1}{2}},\,\textbf{S}^{\sminus}_j \!=( 2S \sminus\, b^\dagger_j b_j)^{\frac{1}{2}}b_j, \\
\textbf{S}^z_l\!=&\sminus\,S+c^\dagger_l c_l,~
\textbf{S}^+_l \!=\!c^\dagger_l (2S \sminus\, c^\dagger_l c_l)^{\frac{1}{2}},~\textbf{S}^{\sminus}_l \!= ( 2S \sminus\, c^\dagger_l c_l)^{\frac{1}{2}}c_l,
\end{split}
\end{align}
meant for anti-ferromagnetic couplings, to study the magnon excitations supported by the ferrimagnetically ordered state. Here, $a_i (a^\dagger_i)$, $b_j^\dagger (b_j^\dagger)$  and $c_l^\dagger (c_l^\dagger)$ are the magnon annihilation (creation) operators for the spins at sites $i \in A$, $j\in B$, $l\in C$ respectively. Subsequently, we write down the magnon Hamiltonian  for the system
\begin{equation}
H =\frac{1}{2} \sum_{{\k}\in BZ} \Psi^\dagger({\k})\; {\cal H}({\k}) \;\Psi({\k})
\end{equation}
 using momentum-space operators $\Psi({\k})\equiv [a^\dagger_{\k}~ b^\dagger_{\k}~c^\dagger_{\k}~a_{\sminus\k}~b_{{\sminus}\k}~c_{{\sminus}\k}]$, where  $\k $ represents Bloch momenta. The operators, $a_{\k}$, $b_{\k}$, $c_{\k}$ etc., have similar definitions as described near \eqn{eq:Fourierc} for the kagome-lattice.  
The $6\times 6$ coefficient matrix $\mcH(\k)$ has the BdG form
\begin{equation}
\!\!\!{\cal H}({\k})=\!\begin{pmatrix}
               ~d_A & ~~~0& ~~~0& ~~~0 &~\gamma_{b\k} & ~\gamma_{c\k}\\
               ~~0   & ~~d_B&~~~0& ~\gamma_{b\k}^*&~~~0&~~~0\\
               ~~0   &~~~0& ~~~d_C&~ \gamma_{c\k}^*&~~~0&~~~0\\
               ~~0&~\gamma_{c\k}&~~\gamma_{c\k}&~~d_A&~~~0&~~~0\\
        \gamma^*_{b\k}&~ ~~0&~~~0&~~~0& ~~d_B&~~~0\\
        \gamma^*_{c\k}&~ ~~0&~~~0&~~~0&~ ~~0&~~d_C\\
               \end{pmatrix}  \label{eq:dice_coe_ham}
\end{equation}
where the diagonal couplings are defined as
$d_A=\, \sum_\eta J_\eta+2\kappa$,
$d_B=  \,J_1+J_3+J_5+2\kappa$, and $d_C= \;J_2+J_4+J_6+2\kappa$, and the non-diagonal couplings describing bonds connecting $A$-$B$ and $A$-$C$ sites are defined as
\begin{align}
\begin{split}
\!\!\!\gamma_{b\k}=& \,(J_1e^{i \frac{k_x}{2}} + J_3e^{-i \frac{k_x}{2}}) e^{i \frac{k_y}{\sqrt{3}}}+J_5e^{-2i \frac{k_y}{\sqrt{3}}},\\
\gamma_{c\k}=& \,J_2e^{2i \frac{k_y}{\sqrt{3}}}+(J_4e^{-i \frac{k_x}{2}} + J_6e^{i \frac{k_x}{2}}) e^{-i \frac{k_y}{\sqrt{3}}},
\end{split}\label{eq:dice_gbgC}
\end{align}
respectively. 
We diagonalize $\mcH(\k)$ (\ref{eq:dice_coe_ham}) numerically (using para-unitary matrices) to obtain three magnon modes that are physical and the rest are BdG copies. In \fig{fig:Dice_lattice}, we show the typical magnon band structure obtained via diagonalization when all the nearest neighbour couplings $J_{1-6}$ are equal.
In this limit, the dice lattice has several symmetries, including $C_6$ rotational symmetry, reflection symmetry about the \emph{six} axes passing through $A$ sites (indicated as dashed lines in \fig{fig:Dice_lattice}(a)), as well as, symmetry under interchange of $B$ and $C$ sublattice sites.
{The symmetries also cause the lower two magnon bands to be degenerate.}

Furthermore, due to the presence of these symmetries, there will be no magnon Hall current for the un-distorted dice lattice. 
Therefore, to obtain a finite Hall response we need to break one or a combination of the above symmetries by introducing distortions through strain or by removing bonds of the dice lattice.
Since, the parameter space of the Hamiltonian in \eqn{eq:Ham_dice} described by the couplings $J_{1-6}$ is 6-dimensional, there are a huge number of possibilities to introduce distortions. However, to be systematic we take the symmetry-breaking route to introduce the deformations in the system. Thus, starting with the most symmetric case, i.e., when all the six couplings are the same, we reduce the above symmetries one by one and obtain various limits of the dice-lattice such as $\aTIII$ lattice, the coupled dumble chain lattice and the boat lattice.\\

\begin{figure}\centering
\includegraphics[width=0.48\textwidth]{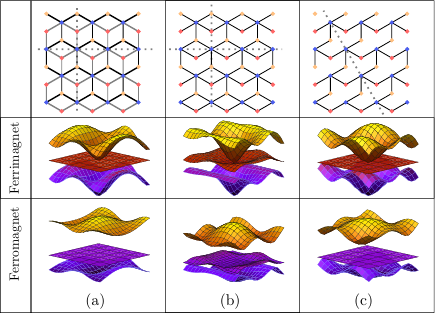}
\caption{Limits of the dice lattice: (a), (b) and (c) represent the three non-trivial lattice limits of the dice lattice, namely the $\alpha\minus{\cal T}^3$ lattice, the coupled dumble chain (CDC) lattice and  the boat lattice (BL), respectively. The dashed line indicates the line of reflection symmetry for each respective lattice. The typical band structures for all three lattices, in both ferrimagnetic and ferromagnetic orders, are also shown.}\label{fig:dice_limits}
\end{figure}

\paragraph{{$\aTIII$} lattice} \label{Sec:Merged_HC_model}
Breaking $B$ and $C$ sublattice interchange symmetry while maintaining $J_1=J_3=J_5$ along with $J_2=J_4=J_6$, reduces the six-fold rotational symmetry $C_6$ to a three-fold rotational symmetry $C_3$ and leaves reflection symmetry about only 3 axes intact (see \fig{fig:Dice_lattice}).
One way to break the interchange symmetry (and more) can be broken systematically is by using a known anisotropy parameter $\alpha$ by setting $J_1/J_2=J_3/J_4=J_5/J_6=\alpha$; hence arriving at the $\aTIII$ lattice \cite{illes2017klein} (see \fig{fig:dice_limits}(a)).
In this way, we can visualize the $\aTIII$ lattice as two identical honeycomb lattices (with or without distortions) superimposed using the tuning parameter $\alpha$. 
For $\!\alpha=0$ ($\infty$), the lattice reduces to the honeycomb limit studied in \Sec{Sec:HC}  with an isolated spin $C$ ($B$) in each unit cell oriented along the easy axis. 
When $\alpha=1$ and $J_1=J_3\neq J_5$, we get two distorted HC lattices merged in equal proportions, with $C_3$ rotational symmetry removed and reflection about the vertical and horizontal axes retained. Typical magnon-band structures for this case, due to ferri and ferro-magnetic orders, are given in \fig{fig:dice_limits}(a) and shows band touching points which are a consequence of the mentioned reflection symmetries.
Setting back $J_1=J_3=J_5$ while keeping $\alpha=1$, we recover the pristine dice lattice with all the original symmetries restored.

In the results section, we study the BC dipoles for this lattice as a function of $\alpha$ for various values of couplings $J_1$, $J_3$ and $J_5$.
\\
\paragraph{Coupled dumble chains}\label{Sec:dumble_model}
Another way to break rotational symmetry is by removing one or more bonds instead of applying strain. 
Starting from an undistorted dice lattice we gradually remove one of the three  $A$-$C$ or $A$-$B$ bonds. For e.g., we vary $J_2$ bond as $\delta J$, tuning $\delta$ from $1\to0$ and keeping the other bonds at equal strength $J$, to approach a lattice consisting of connected chains of dumbles. We name this lattice the coupled dumble chain lattice (see \fig{fig:dice_limits} (b)) and show the reflection symmetries of this lattice about the $x$-axis and the $y$-axis, inherited from the dice lattice, using dashed lines. A typical band structure for $\alpha=0.5$ encountered while approaching the coupled dumble chain lattice from the pristine dice lattice is presented in \fig{fig:dice_limits}(b) and shows no band degeneracies.\\

\paragraph{Boat Lattice}\label{Sec:Boat_lattice_model}
We further gradually remove the $J_3$ bond from the coupled dumble chain lattice, by setting $J_3= \gamma J$ and varying $\gamma$ from $1\to0$, to arrive at another interesting limit -- the boat lattice shown in \fig{fig:dice_limits} (c). Interestingly, the boat lattice is made out of one of the six prototiles in Penrose's $P_1$ tilling \cite[p.~531]{grunbaum1987tilings}  and has only one symmetry (other than lattice translation) which is a reflection symmetry about a slanted axis shown by a dashed line in \fig{fig:dice_limits} (c). As a result, the band structure for the boat lattice (i.e., $\gamma=0$ limit) in \fig{fig:dice_limits}(c) shows a degeneracy between the lower two bands at four $\k$-points of the BZ.\\

Along with the $\aTIII$ lattice, we report the evolution of BCDs obtained by deforming the Dice lattice into the Coupled dumble chain lattice (by varying $\delta$) and then to the Boat lattice (by tuning $\gamma$) in the results section.
The above three lattices are the only non-trivial limits of the dice lattice. Any further removal of bonds either results in an increased unit cell or leads to decoupled structures.  
\subsubsection{Dice Ferromagnet}
The ferromagnetic order on the Dice lattice can be obtained by  making the Heisenberg couplings $J_{1-6}<0$ in the spin Hamiltonian \eqn{eq:Ham_dice}. From which, the {magnon} Hamiltonian for the ferromagnetic order  {is} obtained using the ferromagnetic HP transformations similar to the  transformations used for the  kagome ferromagnet followed by Fourier transformations to the momentum space.  The resulting magnon Hamiltonian will be of the same form as \eqn{eq:kagome_magnon_ham}, i.e.,
\begin{equation}
H =  \sum_{\k\,\in \text{BZ}} \Phi^\dagger({\k})\; {\cal H}({\k}) \;\Phi({\k}),
\end{equation}
where, $\k$ is Bloch momentum and the vector of operators $\Phi({\k})^\dagger\equiv (a^\dagger_{\k}~ b^\dagger_{\k}~c^\dagger_{\k} )$. The momentum space operators $a^\dagger_{\k} (a_{\k}), ~ b^\dagger_{\k}(b_{\k}),~c^\dagger_{\k}(c_{\k})$ are same as  those defined for the dice-ferrimagnet near \eqn{eq:kagome_magnon_ham} and the coefficient Hamiltonian ${\cal H}({\k})$ is a $3\times 3$ matrix given by  
\begin{equation}
{\cal H}({\k})=\begin{pmatrix}
               ~d_A & -\gamma_{b\k} & -\gamma_{c\k}\\
        -\gamma^*_{b\k}&  ~~d_B&~~~0\\
    -\gamma^*_{c\k}&  ~~~0&~~d_C\\
               \end{pmatrix}.
\end{equation}
The symbols $d_A, d_B,d_C, \gamma_{b\k},\gamma_{c\k}$ have the same definitions as in the dice ferrimagnetic case (see \eqn{eq:dice_gbgC}). We diagonalize ${\cal H}({\k})$ numerically using unitary transformations and plot the magnon bands in \fig{fig:Dice_lattice}. When all $J_{1-6}$ are equal, the ferromagnetic magnon Hamiltonian exhibits the symmetries of the pristine dice lattice, including sixfold rotational symmetry $C_6$, reflection symmetry about vertical and horizontal axis as discussed in Ferrimagnetic case in \Sec{Sec:dice_FE}.

Similar to ferrimagnetic order, we study the two BCDs $\Do$, $\Df$ and $\Dex$ on $\alpha\minus{\cal T}^3$ lattice, the coupled dumble chain lattice and the boat lattice for ferromagnetic order ground states using parameters $\alpha, \delta,\gamma $ (as defined for ferrimagnetic case) in the results section.
 \section{Results}
\label{sec:Results}
 In \Sec{sec:BTE}, we have established the magnon spin hall current in terms of Berry curvature dipoles which contain the information about the geometry of the lattice and hence the physical properties of real materials. Now in this section, we provide the results for  the two previously unreported BCDs --$D^{(0)}_{xy}$, $D^{(1)}_{xy}$  and the EBCD, $D^{(\text{ext})}_{xy}$  for various models introduced in \Sec{sec:Models}. The respective magnon currents $j_{\mu}$, $ j_{\text{\tiny bl}}$ and $j_{\text{\tiny SNE}}$ arising from the dipoles $D^{(0)}_{xy}$, $D^{(1)}_{xy}$  and  $D^{(\text{ext})}_{xy}$ can be calculated by scaling the results with the factor $\tau/\hbar V$ ( $V$, $\tau$ defined near \eqn{eq:jy_orig} and \eqn{eq:bte} respectively) and multiplying with  $T,\mu$-gradients.     
\subsection{Honeycomb Lattice}\label{Sec:results_HC}
\subsubsection{Honeycomb antiferromagnet}\label{Sec:results_HC_AF}
\begin{figure}[t]
    \centering
\includegraphics[scale=1.04]{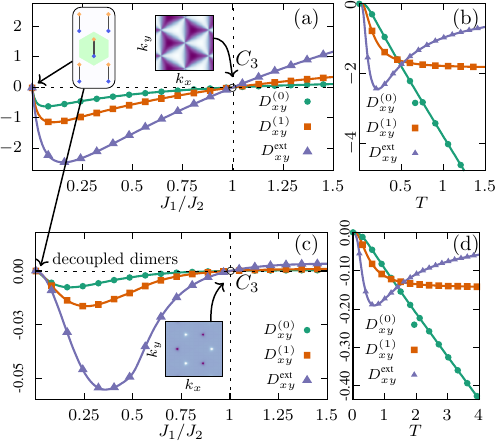}
     \caption{BCDs and EBCD for honeycomb (HC) lattice: (a) The two BCDs $\Do$, $\Df$, and EBCD $\Dex$ are plotted versus $J_1$ (coupling strength of slanted bonds, see \fig{fig:honeycomb}) for antiferromagnetic (AF) order with $J_2 = 1.0$ (coupling strength of vertical bonds), easy-axis anisotropy $\kappa = 0.01$, and temperature $T = 0.2$. (b) The temperature dependence of $\Do$, $\Df$, and $\Dex$ for the AF order with $J_1 = 0.1$, $J_2 = 1.0$, and $\kappa = 0.01$. In (c) we have plotted the three dipoles versus $J_1$ for ferromagnetic (FM) order with $J_2 = 1$, anisotropy difference $\kappa_A - \kappa_B = 0.1$, and temperature $T = 0.2$. (d) The temperature dependence of $\Do$, $\Df$, and $\Dex$ for the FM order with $J_1 = 0.4$, $J_2 = 1.0$, and $\kappa_A - \kappa_B = 0.1$. The two insets in (a) and (c) show the BC distribution ($\Omega(\k)$) for the pristine HC lattice ($J_1=J_2$) for their respective magnetic orders.}
    \label{fig:HC_AF_results}
\end{figure}
We plot the BCDs $\Do$, $\Df$ and EBCD $\Dex$ (defined by \eqn{eq:HC_dipoles}) for the honeycomb anti-ferromagnet versus the coupling strength $J_1$ of the slanted bonds,  in \fig{fig:HC_AF_results}(a), for $J_2=1$ (strength of vertical bonds), $\kappa=0.01$ (easy-six anisotropy) and temperature $k_BT=0.20 $. As a first check, we find in the limit $J_1\rightarrow 0$, the honeycomb lattice reduces to a set of decoupled dimers (see left inset \fig{fig:HC_AF_results}(a)), and therefore as expected the three dipoles in this limit approach zero. When $J_1$ is increased from zero and approaches $J_2$, we encounter a special point at $J_1=J_2$ where the honeycomb lattice is un-distorted possessing full $C_3$ rotational symmetry due to which all the dipoles vanish once again . The $C_3$ symmetry is also manifest in the BC distribution, $\Omega(\k)$ (defined by \eqn{eq:HC_AF_BC}) plotted over the BZ shown as an inset in \fig{fig:HC_AF_results}(a). 

Since all the three BCDs become zero at $J_1=0$ and at $J_1=J_2$, albeit for the distinct reasons given above, a maximum in the BCDs appear between these two values for $J_1$. Beyond $J_1=J_2$, the signature of the three dipole change from -ve to +ve. For $J_1\gg J_2$, the lattice reduces to a set of decoupled chains of spin. Furthermore, due to the equal strain on both slanted bonds, the integral of the $y$-direction BCD distribution ($\partial_{k_y}\Omega$) over the BZ is identically zero irrespective of the value of $J_1$. 
We now turn to the temperature dependence of the dipoles and provide plots of all the three as a function of temperature  in  \Fig{fig:HC_AF_results}(b).  At low temperatures, each of the dipoles approach zero, because at sufficiently low temperatures there are no magnons in the system. As the temperature increases, $\Do$ and $\Df$ increase monotonically, while  $D^{(ext)}_{xy}$ first increases, attains a maximum and then falls to zero  asymptotically. 

\subsubsection{Honeycomb  ferromagnet}\label{Sec:results_HC_F}
Now we move to the HC ferromagnet and  plot $\Do$, $\Df$ and $\Dex$ versus  the coupling  strength of the slanted bonds $J_1$ in \fig{fig:HC_AF_results}(b) with remaining parameters identical to those of the anti-ferromagnetic case, except we set the anisotropies $\kappa_A\neq \kappa_B$, of the two sub-lattices $A$ and $B$ respectively, as well as their difference $(\kappa_A-\kappa_B)=0.1$ to break sub-lattice symmetry (see discussion in \Sec{Sec:model_HC_F} ). The qualitative behavior of the three dipoles, determined by symmetries etc., are similar to those of the antiferromagnetic case, but their values are two orders of magnitude lower compared to those of the anti-ferromagnet. This occurs because, unlike the antiferromagnetic model, the Berry curvature for the ferromagnetic magnons is tuned by the difference in anisotropies $(\kappa_A-\kappa_B)$  (see \eqn{eq:HC_F_BC}) and the maximum BC contribution comes from the vicinity of the two peaks occurring at band touching points located at relatively higher band energies in the BZ. As a consequence, for a given  temperature  on the scale of magnon energy, the contribution of BCDs to magnon-Hall transport has greater chances of being observed in the AFM order than the FM order on the HC lattice. 
To observe BCD-induced transport in the FM ordered HC lattice, materials having significantly higher easy-axis anisotropy difference will be required. To emphasize this point, we have chosen, $(\kappa_A-\kappa_B)=0.1$, a comparatively larger value, for plotting the BCDs in \fig{fig:HC_AF_results}(c, d). The temperature dependence of the three dipoles for the ferromagnet are presented in \fig{fig:HC_AF_results}(d), and shows most of the features to be similar to those of the honeycomb anti-ferromagnet but now occurring at higher temperature scales.

\subsection{Kagome Lattice Ferromagnet}\label{subsec:KLF_results}
\begin{figure}[t]
    \centering\hspace{-10pt}
\includegraphics[scale=1.08]{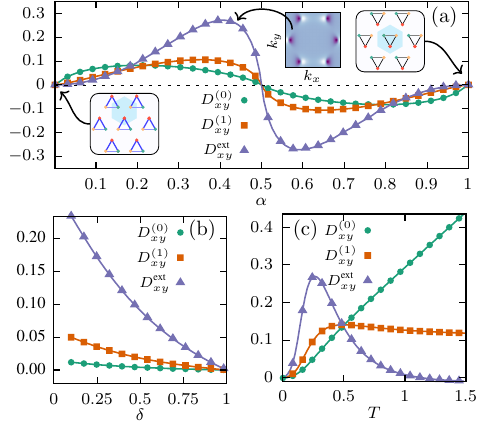}
    \caption{BCDs and EBCD for the kagome lattice: (a) The two BCDs $\Do$, $\Df$, and EBCD $\Dex$ versus $\alpha$ (see \Sec{subsec:KLF_results}) are plotted for ferromagnetic order with strain $\delta=0.75$, easy-axis anisotropy $\kappa=0.01$, and temperature $T=0.25$. (b) shows the variation of the BCDs and EBCD as a function of strain $\delta$ for $\alpha=0.4$ and $T=0.25$, whereas (c) shows the plot of the three dipoles as a function of temperature $T$ for $\alpha=0.4$ and $\delta=0.75$. The left and right insets in (a) show the isolated up and down triangle limits respectively and the middle inset shows BC dipole distribution $\partial_{k_x}\Omega(\k)$ when $\alpha\approx 0.4$.}
    \label{fig:K_F_bc_bcd}
\end{figure}

We study the BCDs of kagome ferromagnet by parameterizing the couplings of the model (see \fig{fig:kagome_ferro} \& \eqn{1.2}) as $J_1=\alpha$ and $J_2=1-\alpha$ by appealing to the symmetry of the lattice.  The value $\alpha=0.5$ corresponds to the symmetric case $J_1=J_2$, whereas in the limit $\alpha\rightarrow 0 (1)$, the lattice transforms into isolated upright (inverted) triangles shown as inset in \fig{fig:K_F_bc_bcd}(a). 
We  plot $\Do$, $\Df$ and $\Dex$ as a function of $\alpha$  for the strain value $\delta=0.75$, easy axis anisotropy $\kappa=0.01$ and temperature $k_B T=0.25$ in  \fig{fig:K_F_bc_bcd} (a). Unlike the HC ferromagnet, the kagome ferromagnet has a significantly large magnitude for the all the three dipoles. Additionally, to have non-vanishing BCDs, the breaking of sub-lattice symmetry by setting unequal easy axis an-isotropies is not required. Breaking of $C_3$ rotation symmetry by setting $\delta\neq 1$ and reflection symmetry by setting  $\alpha\neq 0.5$ is sufficient as discussed in \Sec{Sec:kagome}. 
Due to the reflection symmetry about the horizontal axis shown in \fig{fig:kagome_ferro}, all the three dipoles vanish at $\alpha=0.5$ and are anti-symmetric about this $\alpha$ value.  This anti-symmetry is a consequence of going from  a weakly coupled up-triangle limit to a weakly coupled down-triangle limit, which constitutes a change in the sense of direction since the two limits are flipped versions of each other. As expected, the three dipoles also vanish at the extreme points $\alpha=0$, $1$ corresponding to the disconnected up, down triangle limits, respectively.  We plot the dependence of the three dipoles on strain $\delta$  in \fig{fig:K_F_bc_bcd}(b) for $\alpha=0.2$. It can be seen from the plot, that the magnitude of the three dipoles increases as we increase strain by taking $\delta$ away from {\it unity}. We also plot the temperature dependence of the three dipoles in \fig{fig:K_F_bc_bcd}(c) for $\alpha=0.2$ and $\delta=0.75$. Since the kagome lattice has three bands, the temperature dependence of the dipoles have rich features compared to the honeycomb lattice. The bilinear BCD $\Df$ for the kagome ferromagnet attains  a maximum before saturating to a constant value. Also, the plot of EBCD $\Dex$ on the temperature axis attains a maximum, changes it signature, reaches a minimum (a maximum in other direciton) and then falls off to zero at very large temperatures.  
\subsection{Dice lattice}\label{Sec:reults_dice}
We study the BCDs and EBCD within the three limits of dice lattice (\Sec{Sec:dice}), considering both ferri and ferro magnetic orders,  in the following manner -- {\it 1.} we deform the $\aTIII$ lattice from the honeycomb limit to the dice lattice,  and {\it 2.} we transform the pristine dice lattice to the coupled dumble-chain lattice and subsequently into the boat lattice. 
\subsubsection{$\aTIII$ lattice}\label{Sec:Results_alphat3}
\begin{figure}[t]
    \centering\hspace{-15pt}
\includegraphics[width=0.5\textwidth]{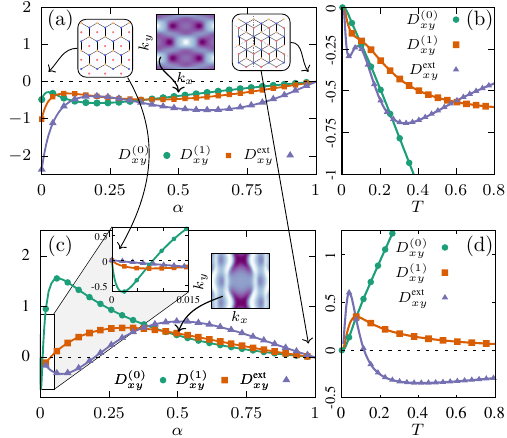}
    \caption{BCDs and EBCD for $\aTIII$ lattice: (a) and (c) show the plots of $\Do$, $\Df$  and $\Dex$ versus $\alpha$ for ferri and ferro-magnetic orders, respectively. We have set other parameters to be $J_1=J_3=0.2$, $J_5=1.0$, ($J_2/J_5=J_4/J_3=J_6/J_1=\alpha$, see \fig{fig:Dice_lattice}), $\kappa=0.01$ and $T=0.2$. (b) and (d) show the temperature variation of the dipole when $\alpha=0.1$ for ferrimagnetic and ferromagnetic orders, respectively keeping all other parameters to be same. The two insets in (a) and (c) shows the BC dipole distribution in $x$-direction, $\partial_{k_x}\Omega(\k)$, for their respective magnetic orders. The leftmost and rightmost insets in (a) show the two limits of the $\alpha T_3$ lattice-- a honeycomb lattice with an isolated site ($\alpha=0$) and the pristine $\alpha T_3$ lattice ($\alpha=1$), respectively.}
    \label{fig:results_alphaTIII}
\end{figure}

\paragraph{Ferrimagnet} As discussed in \Sec{Sec:dice_FE}, the $\aTIII$ lattice combined two identical honeycomb lattices with a control parameter $\alpha$. Therefore, we plot $\Do$, $\Df$ and $\Dex$ versus $\alpha$ in \fig{fig:results_alphaTIII}(a) for the ferri-magnet, in the regime $\alpha\in [0,1)$,  while setting the parameters {$J_1=J_3=0.2$, $J_5=1.0$, ($J_2/J_5=J_4/J_3=J_6/J_1=\alpha$)}, $\kappa=0.01$ and temperature $T=0.2$. We see from the figure at $\alpha=0$, the values of the three dipoles matches with those of HC anti-ferromagnet (HC lattice formed by the $A$ and $B$ sub-lattices, see left inset)  discussed in \Sec{Sec:results_HC_AF} for the same set of parameters. At $\alpha = 1$, the lattice recovers the reflection symmetry about the horizontal axis (see dashed lines in \fig{fig:dice_limits}(a)), resulting in the vanishing of the three dipoles. In this limit, the dice lattice has unequal bond strengths distributed among the six bonds in a manner consistent with the said reflection symmetry, as illustrated in the inset of \fig{fig:results_alphaTIII}(a). 
 In the regime $0< \alpha <1$, the behavior of the dipoles is intersting as it goes through a minimum and maximum before vanishing at $\alpha=1$. For $\alpha > 1$ (not shown), all three dipoles exhibit behavior similar to that for $\alpha < 1$, except with opposite signs. This occurs as they approach another honeycomb (HC) anti-ferromagnet limit formed by the $A$ and $C$ sub-lattices as $\alpha \to \infty$. The temperature dependence of the dipoles is shown in \fig{fig:results_alphaTIII}(b) for $\alpha=0.1$. \\
\paragraph{Ferromagnet}
We plot $\Do$, $\Df$ and $\Dex$ versus $\alpha$ for ferromagnetic order on $\aTIII$-lattice  in \fig{fig:results_alphaTIII}(c) for the couplings $J_1=J_3=0.2$, $J_5=1.0$, ($J_2/J_5=J_4/J_3=J_6/J_1=\alpha$), easy-axis an-isotropy  $\kappa=0.01$ and temperature $T=0.2$. Unlike the HC ferromagnet, unequal values of anisotropies for the three sub-lattices are not required to obtain non-vanishing BCDs and EBCD. Moreover, in the regime $0< \alpha <1$, the magnitudes of the dipoles are significantly larger (relative to both the HC and kagome ferromagnet) and comparable to those of the ferrimagnetic order on $\aTIII$-lattice.  We see from the figure that in the $\alpha=0$ HC limit, the three dipoles go to zero which are consistent with the HC ferro-magnet results  discussed in \Sec{Sec:results_HC_F} for $\kappa_A=\kappa_B$. Even for a small deviation from the HC limit, the three dipoles increase rapidly. Interestingly, this observation also indicates a way to achieve a significant BCD response from the HC ferromagnet, which otherwise exhibits a substantially lower response (see \Sec{Sec:results_HC_F}), by using materials with A-B stacked HC lattice geometries \mycite{mccann2013electronic} and perturbing these materials towards the $\aTIII$ lattice by coupling the A and B layers. Analogous to the ferri-magnetic case, the three dipoles vanish for $\alpha=1$ in accordance with the reflection symmetries discussed in the previous paragraph.  
The temperature plot for the dipoles are presented in \fig{fig:results_alphaTIII}(d) for $\alpha=0.1$. The qualitative nature of the dipoles  is substantially different form ferri-magnet as after a certain temperature, the EBCD $\Dex$ change its sign. 

\subsubsection{ Dice lattice $\to$ coupled dumble chains $\to$ boat lattice}\label{dice_dumble_boat}
We now study the three dipoles as we deform the pristine dice lattice (DL) to the coupled dumble-chain (CDC) lattice  subsequently by the deformation to boat lattice (BL). As discussed in \paras{Sec:dumble_model}{Sec:Boat_lattice_model}, we perform this transformation by following a trajectory in strain-space. First, we vary the strain $\delta$ (\fig{fig:dice_limits}) from $\delta = 1$ to $\delta = 0$, while keeping the other strain $\gamma$ fixed at $\gamma = 1$, to reach the coupled dumble chains lattice. Then, we vary $\gamma$ from $\gamma = 1$ to $\gamma = 0$, while maintaining $\delta = 0$, to arrive at the boat lattice. 
\begin{figure}[t]
\centering\hspace{-12pt}
\includegraphics[width=0.5\textwidth]{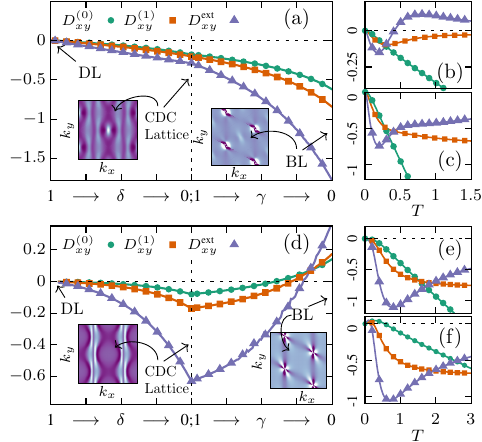}
\caption{BCDs and EBCD evolution while transitioning from {\it dice lattice (DL)} $\to$ {\it coupled dumble chains (CDC)} $\to$ {\it boat lattice (BL)}: (a) and (d) shows the variation of $\Do$, $\Df$ and $\Dex$ over the strain trajectory ($(\delta,\gamma)\in (1,1)\to(0,1)\to(0,0)$) at temperature $T=0.2$ for ferri and ferro-magnetic orders, respectively. The left and right insets show the BC dipole distributions  in $x$-direction, $\partial_{k_x}\Omega(\k)$, for CDC and boat lattice limits, respectively, for their corresponding magnetic orders. The temperature dependence of the dipoles are plotted for the ferrimagnetic order with $(\delta,\gamma) = (0.5, 1.0)$ in (b) and $(\delta,\gamma) = (0.0, 0.5)$ in (c), and for the ferromagnetic order with $(\delta, \gamma) = (0.5, 1.0)$ in (e) and $(\delta, \gamma) = (0.0, 0.25)$ in (f). For all the plots we have set $J=1.0$ and easy-axis anisotropy $\kappa=0.01$.}
    \label{fig:results_dumble_boat}
\end{figure}
\paragraph{Ferrimagnet}
We track $\Do$, $\Df$ and $\Dex$ along the above trajectory for the ferrimagnet and show their behavior as a function of respective strains ($\delta$ or $\gamma$) in \fig{fig:results_dumble_boat}(a). We have set other parameters to be fixed at $J=1$, easy axis anisotorpy $\kappa=0.01$ and  temperature $T=0.2$. As expected, all three dipoles vanish for a pristine dice lattice (\Sec{Sec:dice_FE}) and increase as we apply strain for deforming the lattice towards the CDC-lattice. Next, when we deform the coupled dumble-chain lattice to the boat lattice, along with $D_{xy}$ the dipoles in $y$-direction $D_{yx}$ also become finite. For $\gamma\neq 0,1$, the bond couplings $J$'s are distributed in a manner so as to necessarily produce finite dipoles in two mutually perpendicular direction, and therefore we get  non-zero dipoles in both $x$ and $y$-direction.  It is interesting to observe that boat lattice has non-vanishing dipole even when all bond strengths of the lattice are same. This is because of the non-trivial geometry of the boat lattice (see \fig{fig:dice_limits}(c)) which lacks most of the symmetries of the other lattices we have studied. We show the temperature behavior of the dipoles en route from DL to CDC-lattice (for $\delta=0.5$, $\gamma=1$) and from CDC to BL (for $\delta=0.0$, $\gamma=0.5$) in the leftmost and rightmost insets of \fig{fig:results_dumble_boat}(a), respectively. The typical BCD distribution ($\partial_{k_x} \Omega$) over the BZ in the lowest band for the CDC-lattice and the boat lattice are also shown as insets in the same figure. 

\paragraph{Ferromagnet}
We then track the three dipoles along the strain-space trajectory for ferromagnetic order and plot them as a function of respective strains ($\delta$ or $\gamma$)  in \fig{fig:results_dumble_boat}(b) for temperature $T=0.2$, $J=1$ and easy axis an-isotorpy $\kappa=0.01$. 
While comparing with the ferrimagnetic case, we find the evolution of the dipoles to be similar while transitioning from DL to the CDC-lattice. However, their behavior differs in the transition regime from the CDC-lattice to BL, as in this region the magnitude of the dipoles decrease as we increase strain. The dipoles cross zero at a certain value of $\gamma$, set by temperature, and then increase in the opposite direction. The typical BCD distribution ($\partial_{k_x} \Omega$) over the BZ in the lowest band for both the CDC-lattice and BL are shown as insets in the same figure.
Additionally, we show the temperature behavior of the dipoles transitioning from DL to CDC-lattice (for $\delta=0.5$, $\gamma=1$) and from CDC to BL (for $\delta=0$, $\gamma=0.5$) in the leftmost and rightmost insets respectively.  \section{Summary}\label{sec:summary}
In this article, we provide a proposal to observe previously unreported BC-dipole (BCD) induced magnon-Hall transport in experiments  by simultaneously applying  temperature ($T$)-gradient and spin-injection  to a magnetic insulator (MI).
While the $T$-gradient spatially modulates the temperature profile inside the MI, the spin  injection sets up a spatially varying chemical potential ($\mu$) for the magnons. The confluence of varying $\mu$ and $T$ profiles allows us to  extract previously unaccessed  responses arising from Berry curvature of magnons.We have derived the expressions for the leading order contributions to these two responses. One of which is non-linear in $\mu$-gradient, and the other is a bi-linear response in $\mu$ and $T$-gradients. In addition to the above responses, our arrangement also detects the previously reported response due to extended Berry curvature dipole (EBCD). Therefore, our proposed setup captures the complete picture of geometry-induced magnon transport.
{Subsequently, we apply our expressions for BCD and EBCD to obtain predictions for ferro, anti-ferro, and ferri-magnetic magnons on various experimentally relevant lattice geometries. In the process, we identify salient and interesting features in these models to be observed in experiments.}

In short, our proposal helps us screen topologically trivial magnetic insulators having broken inversion symmetries. Being topologically trivial, these insulators do not show a significant first-order magnon Hall response, as a result they may have been overlooked till now. Despite being topologically uninteresting, a potentially large class of these inversion symmetry broken MIs may possess a non-trivial and rich Berry Curvature band geometry.
We hope that our findings in this paper will significantly broaden the search for such topologically trivial but geometrically rich magnetic materials, and provide a way to characterize these materials that would have been mundane otherwise.

\section*{Acknowledgement}
 AH acknowledges support from DST India via the grant SRG/2023/000118. AR thanks SNBNCBS Kolkata, India for the Bridge fellowship. 
AH thanks Sumilan Banerjee (IISc), Aditya A. Wagh (IISc), Subhro Bhattacharya (ICTS), Anjan Barman (SNBNCBS), and Tanusri Saha-Dasgupta (SNBNCBS) for useful discussions.

 \appendix
 
\section{Magnon chemical potential}\label{ap_chem_pot}

In this appendix, we provide a step-by-step derivation for the magnon chemical potential in the magnetic insulator (the sample) as well as the spin accumulation in the magnetic metals on either side of it. The spin accumulation on the metal at $A$ end of the sample, $\mu_s^A(x)$ by definition satisfies \eqn{eq:js} and \eqn{eq:diffeq-mus}.  The general solution of  diffusion \eqn{eq:diffeq-mus} will be: 
\begin{equation}
    \mu_s^A(x)= \mu_{s-}^A e^{-\frac{x}{l_s}}+\mu_{s+}^A e^{\frac{x}{l_s}}, ~~~~~x\leq -\tfrac{L_x}{2} \nonumber
\end{equation}
Here, $\mu_{s-}^A$ and $\mu_{s+}^A$ are constants of integration. To make the spin accumulation independent of the dimensions of the metal we need to drop the first term in the above equation and therefore we have \begin{equation}
    \mu_s^A(x)= \mu_{s+}^A e^{\frac{x}{l_s}} ,~~~~~~x\leq -\tfrac{L_x}{2}.\nonumber
\end{equation}
we fix the constant $\mu_{s+}^A$ using  \eqn{eq:js} and  write\begin{equation}
    \mu_s^A(x)= -\frac{4e^2 l_s}{\hbar \sigma }\Big( j^{int}_{sA}+\frac{\hbar \sigma_{SH} {\cal E}}{2 e} \Big)e^{\frac{(x+L_x/2)}{l_s}} \end{equation}
Similarly, the spin accumulation in the metal at $B$ end, $\mu_s^B(x)$ (satisfying \eqn{eq:diffeq-mus_B} and \eqn{eq:js_B}) should have the following form
\begin{equation}
    \mu_s^B(x)= \mu_{s-}^B e^{-\frac{x}{l_s}} ,~~~~~~  \frac{L_x}{2}\leq x\nonumber.
\end{equation}
We determine the constant $\mu_{s-}^B$ using \eqn{eq:js_B} and get $\mu_s^B(x)$ in term of interface currents at $B$ joint.
\begin{equation}
    \mu_s^B(x)= \frac{4e^2 l_s}{\hbar \sigma }j^{int}_{sB}\,e^{-\frac{(x-L_x/2)}{l_s}} ,~~~~\frac{L_x}{2}\leq x\nonumber
\end{equation}
The magnon chemical potential  $\mu_m(x)$ in the sample (magnetic insulator) satisfy the diffusion  \eqn{eq:mu_m-diff}, and hence will have a general form as follows:
\begin{equation}
    \mu_m(x)=C_{-} \;e^{-x/l_m}+ C_+\; e^{x/l_m}\label{eq:ap_A_mu_m_gen}
\end{equation}
where $C_\pm$ are arbitrary constants. We can easily re-express $\mu_m(x)$  in the desired form
\begin{equation}
    \mu_m(x)=\frac{\mu_m(\sminus \tfrac{L_x}{2}) \,\sinh \big(\tfrac{L_x-2x}{2l_m}\big)+\mu_m(\tfrac{L_x}{2}) \,\sinh \big(\tfrac{L_x+2x}{2l_m}\big)}{\sinh(L_x/l_m)}\label{eq:ap_A_mu_m_gen}
    \end{equation}
where $\mu_m(\sminus{L_x}/{2})$ and $\mu_m({L_x}/{2})$ are magnon chemical potential at the two interfaces $A$ and $B$ respectively. Now we use boundary conditions \eqref{eq:js_Ax} and \eqref{eq:js_Bx} to express them in terms of interface currents: \begin{equation}
 \mu_m(\sminus \tfrac{L_x}{2})=- 2e \,l_s \frac{ \sigma_{SH} }{\sigma}{\cal E}-j^{int}_{sA}\Big( \frac{\hbar\, \Lambda}{\sigma^{int}_s}+\frac{4 e^2 l_s}{\hbar \sigma}\Big),\label{eq:ap_A_mu_m_A}
\end{equation}
and 
\begin{equation}
 \mu_m( \tfrac{L_x}{2})=\;j^{int}_{sB}\Big( \frac{\hbar\, \Lambda}{\sigma^{int}_s}+\frac{4 e^2 l_s}{\hbar \sigma}\Big)\label{eq:ap_A_mu_m_B}
\end{equation}
We further, use  \eqn{eq:ap_A_mu_m_gen}, in \eqn{eq:js_mag} to get:
\begin{subequations}
\begin{equation}
\resizebox{.48\textwidth}{!}{$\mu_m(\sminus \tfrac{L_x}{2})\coth\big(\tfrac{L_x}{l_m}\big)\sminus\,\mu_m(\tfrac{L_x}{2})\csch\big(\tfrac{L_x}{l_m}\big)=\frac{(j^{int}_{sA}+L_{SSE}\nabla T)}{\hbar\, l_m/\sigma_s}$}\label{muAcondition}
\end{equation}
\begin{equation}
\resizebox{.48\textwidth}{!}{$\mu_m(\sminus \tfrac{L_x}{2})\csch\big(\tfrac{L_x}{l_m}\big)-\mu_m( \tfrac{L_x}{2})\coth\big(\tfrac{L_x}{l_m}\big)=\frac{(j^{int}_{sB}+L_{SSE}\nabla T)}{\hbar\, l_m/\sigma_s}$}\label{muBcondition}
\end{equation}
\end{subequations}
Eliminating $\mu_m(A)$ and $\mu_m(B)$ from  \eqn{eq:ap_A_mu_m_A}, \eqn{eq:ap_A_mu_m_B}, \eqn{muAcondition} and \eqn{muBcondition}  one can express the interface currents, $j^{int}_{sA}$ and $j^{int}_{sB}$  in terms of applied electric field and temperature gredient across the sample. Once it accomplished, one can use \eqn{eq:ap_A_mu_m_A} and \eqn{eq:ap_A_mu_m_B} in \eqn{eq:ap_A_mu_m_gen} to obtain the magnon chemical potential $\mu_m(x)$ in the sample.

\section{Linear  spin wave theory and derivation of bloch Hamiltonian}\label{ap_spin_wv_theo}
In this appendix, we present a generic formalism for deriving the Bloch Hamiltonian for various models introduced in   \Sec{sec:Models} starting from their spin-operator forms in real space. 

We start with anti-ferromagnetic case of honeycomb lattice. The spin Hamiltonian for AF honeycomb, is given by \eqn{eq:HC_Ham} for $\kappa_A=\kappa_B=\kappa$. Using  Holstein-Primakoff (HP) transformation for anti-ferromagnetic order in \eqn{eq:hp_transform_afm}, we can represent Heisenberg interactions between spins $\textbf{S}_i, i \in A$ and  $\textbf{S}_j, j \in B$   in  terms of boson operators up to quadratic order as following:
\begin{align*}
\textbf{S}_i\cdot \textbf{S}_j =\; &(\textbf{S}^+_i \textbf{S}^-_j+\textbf{S}^-_i \textbf{S}^+_j)/2+\textbf{S}^z_i \textbf{S}^z_j\\
=\;& S(a_i b_j+a^\dagger_i b^\dagger_j)-S^2+ S(a^\dagger_i a_i+b^\dagger_j b_j) \end{align*}
and similarly, the easy-axis an-isotropy term
\begin{align*}
\textbf{S}^z_i \textbf{S}^z_i =S^2-2S\,a^\dagger_i a_i , ~~~~\textbf{S}^z_j \textbf{S}^z_j = S^2- 2S\,b^\dagger_j b_j. \end{align*}
We now substitute the above expression in spin Hamiltonian \eqref{eq:HC_Ham} to get
\begin{align}
\begin{aligned}
H\simeq & ~H_{0}- S J_1 \sum_{\langle i,j\rangle\in \text{ slant}} \!\big( a_i b_j+a^\dagger_i b^\dagger_j\big)\\
&~~~~~~- S J_2 \sum_{\langle i,j\rangle\in \text{vertical}}\! \big( a_{i}b_{j}+a^\dagger_{i}b^\dagger_{j}\big)\\
&+S(2J_1+J_3+2\kappa) \Big(\sum_{i} a^\dagger_{i}\,a_{i}+ \sum_{j} b^\dagger_{j}\,b_{j}\Big)
\end{aligned}
\end{align}
Here, $H_{0}= N (2J_1+J_3+2\kappa)S^2$, $N$ being the total number of unit cells, is the  energy of  magnetically ordered ground state and rest part of the right-hand side describes magnon-physics. We note that,  the anti-ferromagnetic Heisenberg interaction results in terms, $a_ib_j$ and $(a^\dagger_ib^\dagger_j)$, that are magnon number non-conserving, much like the case of a superconductor in the theory of electrons. The minimum relative position of spins at $A$ and  $B$ sub-lattices are  $\textbf{r}_{\scriptstyle{/}}= \hat{x}/2 +\hat{y}/(2\sqrt{3})$, $\textbf{r}_{\scriptstyle{\backslash}}= -\hat{x}/2 +\hat{y}/(2\sqrt{3})$ and $\textbf{r}_{\scriptstyle{|}}= \hat{y}/(\sqrt{3})$ . Now we express the magnon Hamiltonian in terms of Fourier space operators (define by \eqn{eq:Fourierab}):
\begin{align*}
\begin{aligned}
H=\sminus  \sum_{\k } \!\big(\gamma_{\k}^* a_{\k}b_{\sminus\k}+\gamma_{\k}a^\dagger_{\k}b^\dagger_{\sminus\k}\big)+ d \sum_{\k} (a^\dagger_{\k}\,a_{\k}+b^\dagger_{\k}\,b_{\k}).
\end{aligned}
\end{align*}
Here, we have defined $d=2J_1+J_3+2\kappa$ and $\gamma_{\k}\equiv\sum_{\eta} J_{\eta} e^{i\k \cdot \textbf{r}_{\eta} }=2J_1 \cos(k_x/2)e^{ik_y/2\sqrt{3}} +J_2 \,e^{-ik_y/\sqrt{3}}$. Two orthogonal momenta are $k_x\equiv \k \cdot \hat{x}$ and $k_y\equiv \k \cdot \hat{y}$. The matrix form of the magnon Hamiltonian is  \eqn{eq:Mag_Ham_HCAF}. The coefficient matrix ${\cal H}_{\k}$ in \eqn{eq:hc_afm_bdg}, satisfy time reversal property ${\cal H}_{\k}={\cal H}^*_{-\k}$.
Moreover,  ${\cal H}_{\k}$ has a Bogoliubov–de Gennes
(BdG) form (see \cite{kondoNonHermiticityTopologicalInvariants2020}),  as it can be written as $ {\cal H}_{\k}=\begin{pmatrix} h_{\k}& \Delta_{\k}\\
\Delta^*_{-\k}&h^*_{-\k} \end{pmatrix}$ where $h_{\k}=\begin{pmatrix} d&0\\
0&d \end{pmatrix}$, $\Delta_{\k}=\begin{pmatrix} 0&\gamma_{\k}\\
\gamma^*_{\k}&0 \end{pmatrix}$ and they satisfy $h_{\k}=h^\dagger_{\k}$ and $\Delta^T_{\k}=\Delta_{-\k}$ . The BdG Hamiltonian are diagonalized using para-unitary matrices (say ${ P}$) satisfying, ${ P}^{-1}\Sigma^z { P}=\Sigma^z$. Other equivalent procedure, to digonalize BdG Hamiltonian is to diagonalize  $\Sigma^zH$ through similarity transformations. The latter method have been used for BdG system studied in this article.

We can obtain the magnon coefficient Hamiltonian for the rest of models using the similar procedure as above. 
\subsection{Berry curvature calculation}
Next, we will briefly discuss the methodology used for calculating the Berry curvature in the various models we studied. It is possible to analytically diagonalize the magnon coefficient Hamiltonian for the honeycomb lattice in both the anti-ferromagnet and ferromagnet cases, see \Sec{app_sec_ana_BC_HC}. For the remaining three models, we numerically diagonalize the magnon coefficient Hamiltonian and use the Wilson loop method (see \Sec{app_sec_Wilson_method}) to calculate their Berry curvature.
\subsubsection{Analytical expression of BC for honeycomb-lattice}\label{app_sec_ana_BC_HC}
\paragraph*{Anti-ferromagnet}
To obtain the spectrum of  BdG Hamiltonian $\mcH_{\k}$ for HC anti-ferromagnet (given in \eqn{eq:hc_afm_bdg}), we need to   diagnalize following matrix;  
\begin{equation}
\Sigma^z\mcH_{\k}=\begin{pmatrix}
 ~~d & ~~0 &~~0&\gamma_{\k\!} \\
~~0&~~d&\gamma^*_{\k}&~~0\\
~~0&-\gamma_{\k}& -d&~~0\\
-\gamma^*_{\k}&~~0&~~0&-d~
\end{pmatrix}
\end{equation}
The diagonal form of $\Sigma^zH$ 
is ${P} \Sigma^z\, \mcH_{\k} { P}^{-1}=\text{dia}(E_1,E_2,-E_1,-E_2)
$ where $E_{1,2}= \sqrt{d^2-|\gamma_{\k}|^2}\nonumber$ and the para-unitary matrices is
\begin{align*}
P=\begin{pmatrix}
\cosh\frac{\beta}{2}& 0 & 0&\sinh\frac{\beta}{2}\,e^{i\phi} \\
0  &\cosh\frac{\beta}{2}& \sinh\frac{\beta}{2}\,e^{-i\phi}&0\\
0  &\sinh\frac{\beta}{2}\,e^{i\phi}&\cosh\frac{\beta}{2}&0\\
\sinh\frac{\beta}{2}\,e^{-i\phi}&0&0&\cosh\frac{\beta}{2}
\end{pmatrix}
\end{align*}
where $\tan \phi= \arg(\gamma_{\k})$ and  $\cosh\beta=d(d^2-|\gamma_{\k}|^2)^{-1/2}$. The first two energies, $E_1, E_2$ are physical magnon bands whereas $-E_1, -E_2$ are their BdG counterparts. Further, the two magnon bands are degenerate i. e. $E_1=E_2$.
The Berry curvature of the four magnon bands can be calculated using following formula prescribed for BdG system\cite{kondoNonHermiticityTopologicalInvariants2020}. 
\begin{align}
\bf{\Omega}(\k)\equiv&\;i\; [\partial_{k_x}({  P}^{-1} \Sigma^z \partial_{k_y}{  P})-(k_x\leftrightarrow k_y)]. \end{align}
Substituting $P$ in above formula, we find that the Berry curvatures of two physical magnon bands is given by $\Omega(\k)$ and $-\Omega(\k)$ where
\begin{align}
\Omega(\k)=-\frac{i}{4}\;\frac{d}{(d^2- |\gamma_{\k}|^2)^{3/2}}\;\frac{\partial (\gamma_{\k}^{}~ \gamma^*_{\k})}{\partial (k_x~k_y)}
\end{align}
It is now obvious that for non-zero Berry curvature one need to have $\gamma_{\k}$ be a complex quantity (neither real nor pure imaginary). We use expression of $\gamma_{\k}$, to get the Berry curvature for honey-comb antiferromagnetic magnon which is $$\Omega(\k)=\frac{dJ_1(J_1\sin(k_x)-2J_2 \sin(k_x/2)\cos(\sqrt{3}k_y/2))}{4\sqrt{3}(d^2-|\gamma_{\k}|^2)^{3/2}}.$$
The $\Omega(\k)$ satisfy the time-reversal property, which will ensure the Chern-numberof the system to be zero i. e. $\int_{\text{\tiny BZ}} \Omega(\k)=0$. The fact that the two slanted bonds have identical bond couplings strengths, $J_1$ will results into the property $\Omega(k_x,-k_y)=\Omega(k_x,k_y)$ and is the cause of dipoles along $y$-direction to be zero for all values of $J_1$ and $J_2$. 
\paragraph*{Ferromagnet} The eigenvalues and eigen-vectors of the Bloch Hamiltonian (\ref{4.B.3}) for honeycomb ferromagnet are
\begin{equation}
E_{\pm}= d_A+d_B\pm ((d_A-d_B)^2+|\gamma_{\k}|^2)^{1/2} 
\end{equation}
and,
\begin{equation}
U=\begin{bmatrix}
\cos\frac{\beta}{2}&-\sin\frac{\beta}{2}\,e^{-i\phi}\\
\sin\frac{\beta}{2}\,e^{i\phi}&\cos\frac{\beta}{2}
\end{bmatrix}\label{eq:eigen_vec_HC_F}
\end{equation}
where $\tan \phi= \arg(\gamma_{\k})$ and  $\cosh\beta=(d_A+d_B)/[(d_A-d_B)^2+|\gamma_{\k}|^2]^{1/2}$. The used formula for BC calculation in the ferromagnetic case is
\begin{align}
\Omega(\k)\equiv&\;i\; [\partial_{k_x}({U}^{-1}  \partial_{k_y}{U})-(k_x\leftrightarrow k_y)]. 
\end{align}
We use \eqn{eq:eigen_vec_HC_F} in the above formula to obtain the required Berry curvautre for honeycomb ferromagnet given in \eqn{eq:HC_F_BC}.
\subsubsection{Numerical BC calculation for kagome and dice lattice models}\label{app_sec_Wilson_method}
For the numerical computation of BC it is necessary to  divide the Brillouin zone (BZ) into discrete points. 
In Wilson loop method  one defines the link variables $ U_{\mu}(\k_r)$  variables associated to link connecting the two adjacent points in the $\mu$-direction, $\k_r$ and $\k_{r}+\delta \k_{\mu}$ in discrete BZ: 
\begin{equation}
    U_{\mu}(\k_r)= \frac{\braket{n(\k_r)}{ {\cal S}| n(\k_{r}+\delta \k_{\mu})}}{|\braket{n(\k_r)}{{\cal S}|n(\k_{r}+\delta \k_{\mu})}|}
\end{equation}
Where ${\cal S}=\Sigma^z$  for BdG systems, ${\cal S}=\cal{I}$ for non-BdG systems and $\ket{n(\k_r)}$ are the eigenvectors of $\mcH_{\k}$ at point $\k_r$. Then Berry curvature along $z$-direction, $\Omega^z$ can be approximated by the flux $F_{12}$ passing through the plaquettes of the discrete BZ;
\begin{equation}
 F_{12}(k_r) \delta\k_1\delta\k_2=\ln\Big[\frac{U_1(\k_r)U_2(\k_{r}+\delta \k_{1})}{U_1(\k_r+\delta \k_{2})U_2(\k_{r})}\Big]
\end{equation}

 \bibliography{ref}
\end{document}